\documentclass[preprint]{aastex61}

\submitjournal{ApJ}

\shorttitle{Quantifying weak non-thermal solar radio emission}
\shortauthors{Sharma et al.}
\newcounter{attnctr} 

\usepackage[normalem]{ulem}

\begin{document}

\title{Quantifying weak non-thermal solar radio emission at low radio frequencies}

\correspondingauthor{Rohit Sharma}
\email{rohit@ncra.tifr.res.in}

\author[0000-0002-0786-7307]{Rohit Sharma}
\affil{National Centre for Radio Astrophysics, Tata Institute of Fundamental Research, Pune 411007, India}

\author{Divya Oberoi}
\affiliation{National Centre for Radio Astrophysics, Tata Institute of Fundamental Research, Pune 411007, India}

\author{M.~Arjunwadkar}
\affiliation{Centre for Modeling and Simulation, Savitribai Phule Pune University, Ganeshkhind, Pune 411007, India}

%% Note that the \and command from previous versions of AASTeX is now
%% depreciated in this version as it is no longer necessary. AASTeX 
%% automatically takes care of all commas and "and"s between authors names.

%% AASTeX 6.1 has the new \collaboration and \nocollaboration commands to
%% provide the collaboration status of a group of authors. These commands 
%% can be used either before or after the list of corresponding authors. The
%% argument for \collaboration is the collaboration identifier. Authors are
%% encouraged to surround collaboration identifiers with ()s. The 
%% \nocollaboration command takes no argument and exists to indicate that
%% the nearby authors are not part of surrounding collaborations.

\begin{abstract}
The recent availability of fine grained high sensitivity data from the new generation low radio frequency instruments such as the Murchison Widefield Array (MWA) have opened up opportunities for using novel techniques for characterizing the nature of solar emission at these frequencies.  
Here we use this opportunity to look for evidence for the presence of weak non-thermal emissions in the 100-240 MHz band, at levels weaker than have usually been probed.  
The presence of such features is believed to be a necessary consequence of nanoflare-based coronal and chromospheric heating theories.  
We separate the calibrated MWA solar dynamic spectra into a slowly varying and an impulsive, and hence non-thermal, component.
We demonstrate that Gaussian mixtures modeling can be used to robustly model the latter and we estimate the flux density distribution as well as the prevalence of impulsive non-thermal emission in the frequency-time plane.
Evidence for presence of non-thermal emission at levels down to $\sim$0.2 SFU (1 SFU = 10$^4$ Jy) is reported, making them the weakest reported emissions of this nature.
Our work shows the fractional occupancy of the non-thermal impulsive emission to lie in the 17--45\% range during a period of medium solar activity.
We also find that the flux density radiated in the impulsive non-thermal emission is very similar in strength to that of the slowly varying component, dominated by the thermal bremsstrahlung.
Such significant prevalence and strength of the weak impulsive non-thermal emission has not been appreciated earlier.
\end{abstract}

\section{Introduction}

Nanoflare based coronal and chromospheric heating proposed by \citet{Parker1988} continues to be regarded as a plausible model for resolving the long-standing coronal heating problem.
Over the past few decades, the bulk of the effort for detecting nanoflares has been concentrated in the X-ray and EUV bands.
Major improvements in instrumental capabilities in these wavebands over this period have enabled tremendous progress in building a much more comprehensive understanding of energetic impulsive events and have allowed researchers to probe ever weaker events and emissions.
It has also been established that in order for nanoflare based heating models to be effective, the power-law distribution of the total flare energy, $W$, ($dN/dW \propto W^{\alpha}$) must have a slope, $\alpha \leq -2$ \citep{Hudson1991}.
Though researchers have been able to detect weaker flares, the energy of the most studied bursts lie in the range 10$^{26}$--10$^{31}$ ergs \citep{hannah2011}. 
Most studies in the X-ray and EUV bands however have found $\alpha > -2$.
In the X-ray band \citet{hannah2011} have shown that the intermittancy of these flares, their distribution on the solar surface and association with active regions, all argue that these events do not provide an adequate explanation for coronal heating.
The same arguments apply to the EUV flares as well.
The quest to look for even weaker impulsive emissions with characteristics meeting the requirements for coronal heating continues.

Radio bursts of different types have been studied for a long time and their distributions have been characterized in detail \citep[e.g.][]{McLean1985}.
Impulsive radio emission provides direct evidence for the presence of non-thermal electrons. 
In the present context, these radio emissions are believed to be produced by electron beams, likely energized by small reconnection events, travelling along magnetic field lines.
The emission mechanisms involved are expected to be plasma emissions, which give rise to coherent emission at the local plasma frequency and its harmonic.
This coherent nature of radio emission offers the significant advantage, over the thermal nature of flare associated emission in the X-ray and EUV bands, that even energetically weak events can give rise to large observational signatures.
Energetically, the weakest reported impulsive events have indeed been at low radio frequencies; \citet{Ramesh2013} reported type-I radio bursts with energies in the range 10$^{21}$ erg.
Though the uncertainty in the associated energy estimate is rather large, such bursts are still by far the weakest reported.
Due to their intermittancy and associations with other energetic events on the Sun these known bursts also do not meet the requirements for coronal heating.

New generation of low radio frequency arrays have recently become available.
Examples include the Long Wavelength Array \citep[LWA;][]{Kassim2010} in the USA, LOw Frequency ARray \citep[LOFAR;][]{Haarlem2013} in the Netherlands and the Murchison Widefield Array \citep[MWA;][]{Tingay2013-MWA-design} in Australia.
These more sensitive instruments provide fine-grained data over wide bandwidths and are already enabling new and interesting studies \citep[e.g.][and many others]{Oberoi2011,Morosan2014,TunBeltran2015}.
The data from these new instruments provides fertile new ground to explore for presence of weak non-thermal emissions and their suitability for coronal heating.
In a previous work we used a wavelet based technique to identify and characterise the non-thermal emission features seen in the MWA dynamic spectra (DS) \citep{Suresh2017}.
These numerous short-lived (1--2 s) narrow-band (4--5 MHz) non-thermal emission features were also some of the weakest reported instances of non-thermal radio emission.
Here we use a different method to characterize similar non-thermal radio emissions.
The previous approach relied on being able to robustly identify emission features in the calibrated DS, while the present one is based on the analysis of flux density distribution of the DS.
We are motivated, in part, by the expectation that not having the requirement for these emission features to be identified as such in the DS will allow us to probe even weaker instances of non-thermal emission.
It also allows to quantify aspects of non-thermal emission not addressed by the previous work.

Section \ref{ins_obs} describes the relevant capabilities of the MWA and the data used here along with the state of the Sun during these observations.
Some aspects of solar radio emission relevant for this work are discussed in Sec. \ref{Sec:solar-emission}.
Our implementation of the Gaussian mixture method to quantify the presence of weak non-thermal solar radio emissions and the results obtained from this analysis are described in Sec. \ref{Sec:gm}; while Sec. \ref{Sec:uncert-estimates} provides an assessment of the uncertainty associated with these results.
A discussion of these results is presented in section \ref{result}, followed by conclusions in Sec. \ref{conclusion}.
\section{Observations and Pre-processing}\label{ins_obs}
Located in the very radio quiet Western Australia, the MWA is the low frequency pre-cursor for the Square Kilometre Array (SKA).
Its elements comprise 16 dual polarization dipoles arranged in a 4$\times$4 configuration, referred to as tiles, and cover the radio band from 80 to 300 MHz.
Details of the MWA design are available in \citet{Lonsdale2009-MWA-design} and \citet{Tingay2013-MWA-design}; and an overview of its key science targets, including solar, heliospheric and ionospheric science, is available in \citep{Bowman2013-MWA-Science}.

\begin{figure}%[htbp]
\begin{center}
\begin{tabular}{ccc}
\includegraphics[trim={0.0cm 0cm 0.0cm 0.0cm},clip,scale=.13]{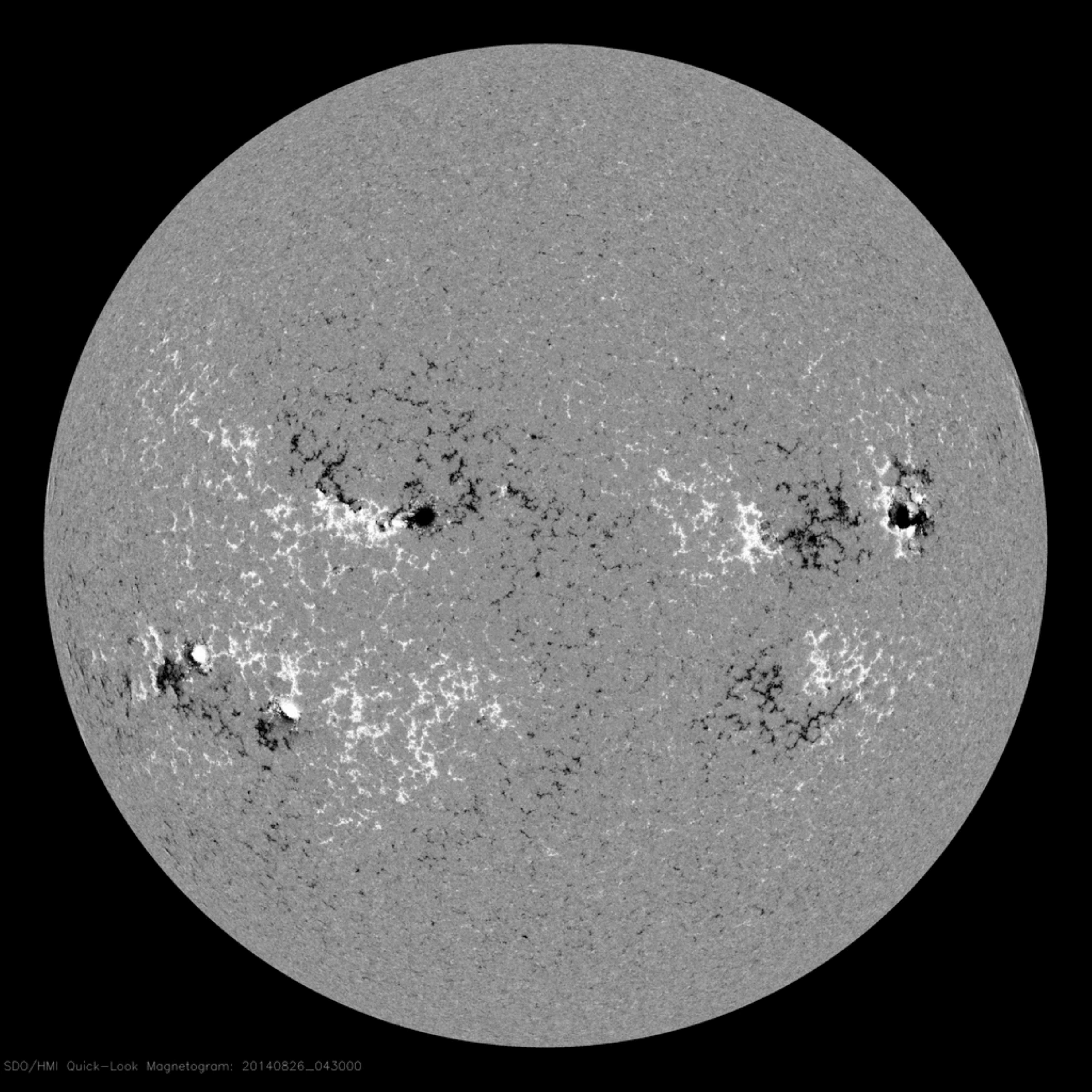}
 &
\includegraphics[trim={1.5cm 0cm 1.5cm 0.0cm},clip,scale=0.13]{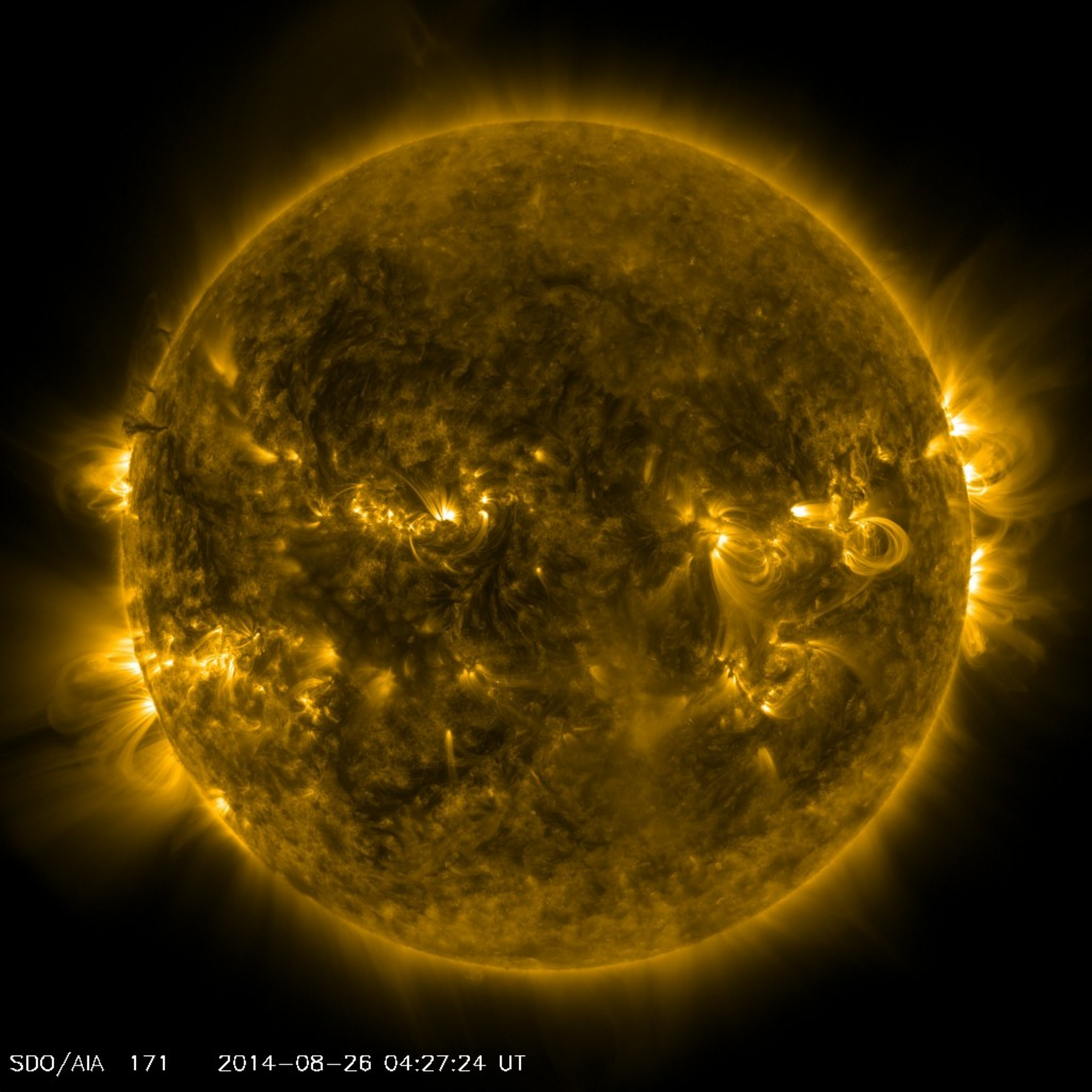}
 &
\includegraphics[trim={0.0cm 0cm 0.0cm 0.0cm},clip,scale=0.33]{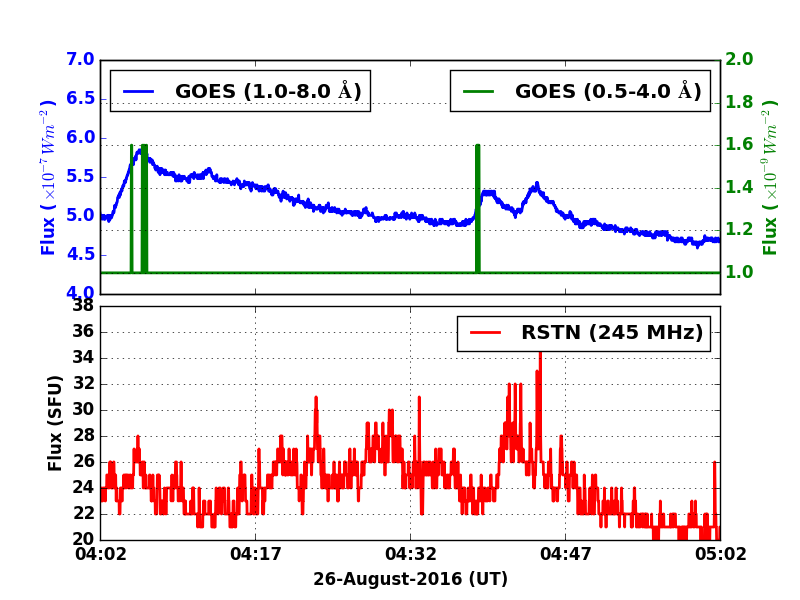}\\
\end{tabular}\\
\end{center}
\caption{The left panel shows the HMI magnetogram taken at 04:30 UT and the middle panel the AIA 171 $\AA$ image taken at 04:27 UT. 
The right panel shows the soft X-ray (in blue), hard X-ray (in green) and 245 MHz radio flux timeseries of the Sun taken from GOES and RSTN instruments respectively. 
\label{Fig:Sun-general}}
\end{figure}

The MWA data presented here were taken on August 26, 2014 from 04:02 to 05:02 UT, as a part of the solar observing proposal G0002.
The level of solar activity on this day is described as medium by {\tt solarmonitor.org}. 
Eight NOAA active regions could be seen on the visible solar disk, and 3 C-class X-ray flares were reported, none of them in our observing period. 
The NOAA Space Weather Prediction Center lists 5 significant and 1 minor group of radio type-III bursts along with 8 minor and 4 significant isolated type-III radio bursts for this day.
Figure \ref{Fig:Sun-general} shows a magnetogram from the Helioseismeic Magnetic Imager (HMI) and a 171 $\AA$ image from the Atmospheric Imaging Assembly (AIA), both on-board Solar Dynamic Observatory taken during our observing interval.
It also shows the GOES soft X-ray, hard X-ray and Radio Solar Telescope Network (RSTN) 245 MHz time series for the duration of our observations.
In this period, the GOES soft X-ray flux lies in the B class flare range and varies between $\sim$5--6$\times 10^{-7}$ $W\ m^{-2}$ and the hard X-ray flux varies in the range from 1.0--1.6 $\times 10^{-9}$ $W\ m^{-2}$.
Though the variations in the observed soft X-ray flux are small, a few weak peaks can distinctly be seen. 
The only radio event reported in our observing interval at 04:41 UT appears to arise due to radio frequency interference, and has been flagged in Fig.\ \ref{Fig:Sun-general}. 

%%----------------------------------------------------------------------------------

%%----------------------------------------------------------------------

The MWA architecture allows one to distribute the 30.72 MHz of observing bandwidth flexibly across its entire observing bandwidth using 24 {\em coarse} spectral channels, each 1.28 MHz wide.
Each coarse channel comprises of 32 {\em fine} spectral channels of 40 kHz each.
Utilizing this flexibility, these observations were carried out in the so-called {\em picket fence} mode, with observing bandwidth distributed in 12 groups of 2 contiguous coarse channels each, distributed in a roughly log spaced manner between 80 and 240 MHz.
The time and frequency resolution of these data are 0.5 s and 40 kHz, respectively.
Here we only present the analysis for the XX polarization at the nine frequencies higher than 100 MHz for six baselines.
These observing bands were centered at 109.44, 120.96, 133.76, 146.56, 161.92, 179.84, 197.76, 218.24 and 241.23 MHz.
The analysis for the YY polarization is analogous. Six baselines used for analysis are formed by combinations of Tile011, Tile021, Tile022 and Tile023.

The MWA dynamic spectra observed on individual short baselines are flux calibrated following the prescription given in \citet{Oberoi2017a}.
Briefly, this technique relies on a mix of laboratory and field measurements to estimate the contributions to the system temperature from the receiver electronics and the ground pickup; detailed modeling to estimate the primary beams of MWA tiles; and using baselines short enough to keep the Sun unresolved (3.5--26.0$\lambda$).
The contribution of the sky is estimated using the all-sky brightness temperatures maps at 408 MHz \citep{Haslam1982-408-map}, scaled to the observing frequency using a spectral index of 2.55 \citep{Guzman2011-Tsky-spectral-index}.
The flux calibrated dynamic spectra produced following this prescription form the starting point for the analysis presented here.
The top panel of Figure \ref{Fig:flux} shows the observed dynamic spectrum in Solar Flux Units or SFU (1 SFU = 10$^4 Jy$) for all nine frequency bands for the baseline Tile011-Tile022. 

%%--------------------------------------------------------

\section{Nature of solar radio emission}\label{Sec:solar-emission}
%{\bf Overall general description, including the relationship between impulsive and non-thermal emission and advantages of the low radio frequency regime.}
During periods of low solar activity, the bulk of the low frequency radio solar emission corresponds to the thermal emission from the million K corona.
The optical depth, $\tau$, of the corona varies with frequency, leading to the change in the observed mean solar brightness temperature, $T_{\odot}$, with frequency \citep{McLean1985}. 
The thermal emission is spectrally smooth and is expected to evolve slowly, on time scales consistent with large scale changes in the corona. 
In addition, the Sun also has emission components which evolve on much faster time and shorter spectral scales.
The well known solar bursts type-I through V are examples of such emissions and have been characterized in great detail in the literature \citep[e.g.][]{McLean-Sheridan-1985}.
With the availability of newer generation of instruments, there is also an increasing awareness of the presence of numerous comparatively weak, narrow-band short-lived emission features \citep[e.g.][]{Oberoi2011,Morosan2015,TunBeltran2015,Suresh2017}, which must be non-thermal in origin.
There is a possibility that these weak emissions correspond to the radio signatures of \citet{Parker1988} nanoflare like events.
As discussed earlier, the coherent nature of emission mechanisms involved at low radio frequencies make this a primary part of band to look for such emissions.
%Unlike X-rays and EUV wavelengths, where the emission associated with nanoflares is thermal, the emission at low radio frequencies arises from coherent emission mechanisms, implying that even energetically weak events can give rise to large observational signatures, making these frequencies a promising place to look for their signatures.

%{\bf Separating the observed data into a thermal and non-thermal parts by separating it into slowly varying and impulsive features. Intepretation of these histograms}
Our objective, in this non-imaging study, is to quantify the presence of radio impulsive features observed, especially the weaker ones and assess their suitability for contributing to coronal heating.
Towards this end, we model the observed dynamic spectra as a superposition of slowly varying and impulsive components.
As argued above, the former can be expected to be dominated by the thermal emission, and the latter by the non-thermal emission.
It can be shown that in the strong signal regime, applicable for these solar observations, the probability distribution of amplitudes of measured interferometric cross-correlations (visibilities) follows a Gaussian distribution (Appendix A).
An interferometric baseline to which the Sun appears as an unresolved source measures the flux density integrated over the entire solar disc, $S_{\odot}$.
%Irrespective of the inhomogenities in the solar surface brightness distribution \citep[e.g.][]{Mercier-Chambe2012}, a baseline which sees the Sun as an unresolved point source, 
The measured values of thermal contribution to $S_{\odot}$ over a narrow bandwidth and durations short compared to those at which the emission is expected to change are hence expected to follow a Gaussian distribution.
The impulsive contribution to $S_{\odot}$ must then appear at higher flux densities and depending on the nature and distribution of these emissions, the observed $S_{\odot}$ histograms can take shapes ranging from showing a skewness to the right, to a high flux density tail or even clusters of emission at higher flux densities.

\begin{figure}%[htbp]
\centering
\begin{tabular}{c}
\includegraphics[width=0.65\textwidth]{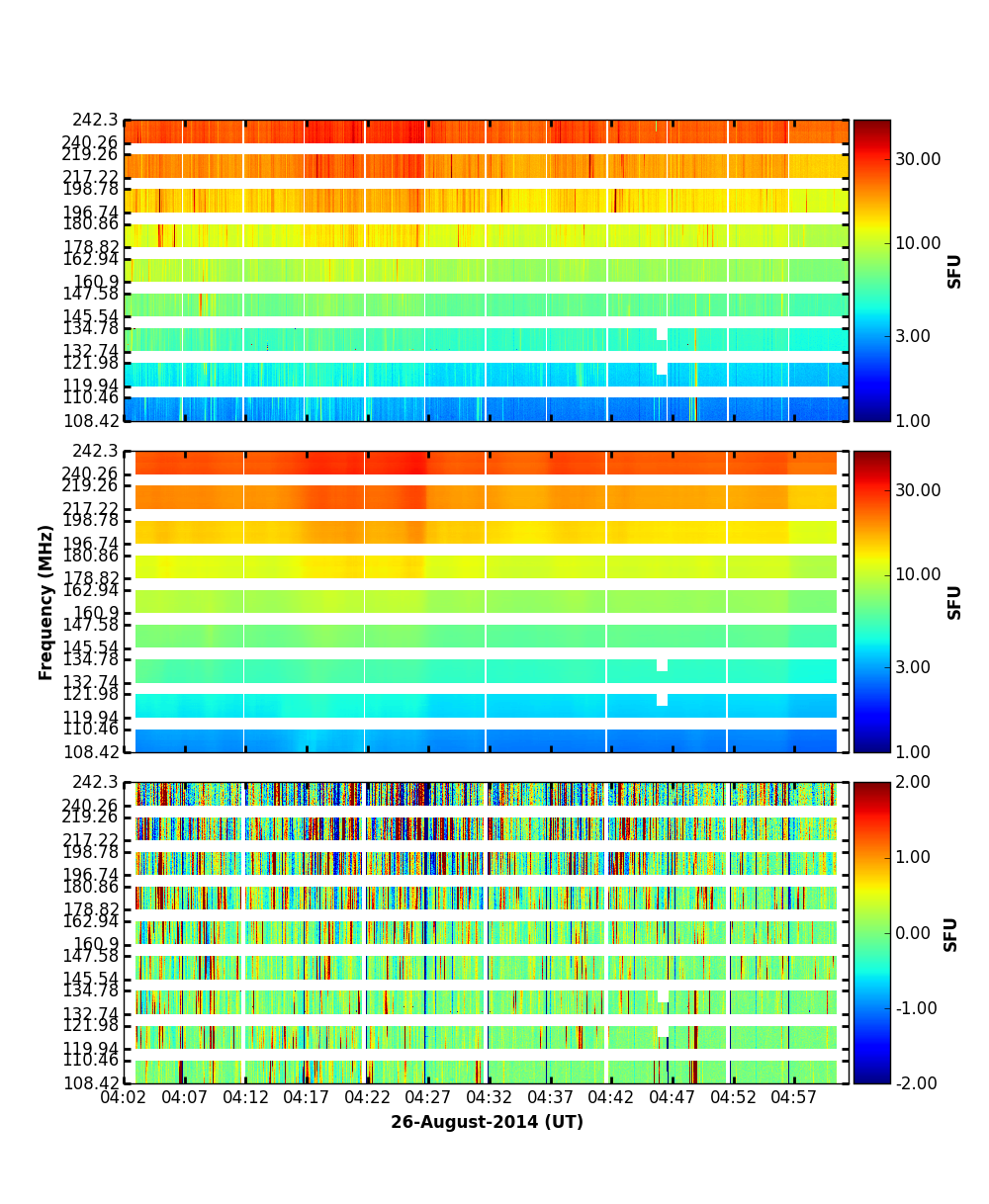}\\
\end{tabular} 
\caption{The observed dynamic spectrum for 26 August 2014 from 04:02 to 05:02 UT is shown in the top panel .
The middle and bottom panels show the corresponding slowly varying and impulsive emission components, respectively.
All fluxes are in SFU.
\label{Fig:flux}}
\end{figure}

%{\bf Details of practical implementation.}
In practice, we find that for the data used here a robust median filter using a window of 120 s is efficiently able to separate the thermal continuum from the impulsive emission. Varying the median filter between 60 and 120 s did not lead to a discernible change in distribution of the filtered DS.
The observed $S_{\odot}$ dynamic spectrum (DS) along with its decomposition into continuum and impulsive emission components are shown in Fig.\ \ref{Fig:flux}.
We find that the underlying thermal continumm is rather steady, usually varying by $<$10 $\%$ in time depending on the frequency with typical variation on time scales of many minutes. 
The impulsive part of the emission is obtained by subtracting the slowly varying DS from the observed DS and is shown in the bottom panel of Fig.\ \ref{Fig:flux}.
To emphasize the presence of numerous weak impulsive features, the color scale on this panel has been saturated at $\pm$2 SFU.
Most of the impulsive features are weaker than a few SFU.
To quantitatively estimate the prevalence, or the fractional occupancy, of these features in the time-frequency plane and this strength we use the Gaussian mixture method, described in Sec. \ref{Sec:gm}.

%%----------------------------------------------------------------------
%%----------------------------------------------------------------------

\section{Gaussian Mixtures Modeling and Results}
\label{Sec:gm}

Gaussian mixtures modeling (GMM) is a probabilistic framework which assume that the data have come from a parent distribution consisting of a superposition of a finite number of Gaussians \citep{Titterington1985,McLachlan2000}. 
Using the data and parametric estimation methods such as the maximum likelihood method, it estimates the parameters of these constituent Gaussians.
As argued in Sec. \ref{Sec:solar-emission}, it is reasonable to expect the multiple measurements of the relatively stationary thermal solar emission to follow a Gaussian distribution and for non-thermal emission features to show up at higher flux density values.
In this work, we have subjected the impulsive DS, obtained by subtracting the slowly varying median filtered DS from the observed DS, to Gaussian mixture analysis.
Our choice is inspired by the fact that, when used with care, GMMs provide a flexible and reasonable approach to model distributions with complex shapes.
As we establish later, GMM is indeed able to provide compact and robust descriptions of the flux density distributions of impulsive DS.
For the data being considered here, which has medium level of activity,
the Gaussian corresponding to this slowly varying thermal emission is expected to have the largest weight and a mean close to zero.
%Though we have no {\em a-priori} expectations for the impulsive emission to follow Gaussian distributions, the data can be described well by a relatively small number of Gaussian components.

In the GMM approach, we model the probability density function $f$ of the data $x$ as the summation of $K$ univariate Guassians

\begin{equation}
f(x;\{w,\mu,\sigma^{2}\},K) = \sum_{i=1}^{K} w_i \phi(x;\mu_i,\sigma_i),
\end{equation}
where $\phi$ is the Gaussian probability density function
\begin{equation}
\phi(x;\mu,\sigma) = \frac{1}{\sqrt{2 \pi \sigma^2}} \exp \Big(-\frac{(x- \mu)^2}{\sigma^2} \Big).
\end{equation}
We obtain maximum-likelihood estimates of the parameters of the constituent Gaussians, namely their means $\mu_i$, weights $w_i$, and standard deviations $\sigma_i$, using the Expectation Maximization (EM) algorithm \citep{McLachlan2000}.
The weight $w_i$ can be also interpreted as the area under the $i$th Gaussian component. 
As the area under the entire distribution is normalized to unity, $w_i$ represents the fraction of data belonging to the $i$th Gaussian component.

We fit GMMs up to $K$ components to the data using the EM algorithm, followed by model selection using the Bayesian Information Criterion (BIC) \citep{Burnham2002}. We set the upper bound $K$ to 15 by trial and error. However, for our data, BIC-based model selection always led to optimal models which have $K < 10$.

\begin{figure}
\centering
\includegraphics[width=0.8\textwidth]{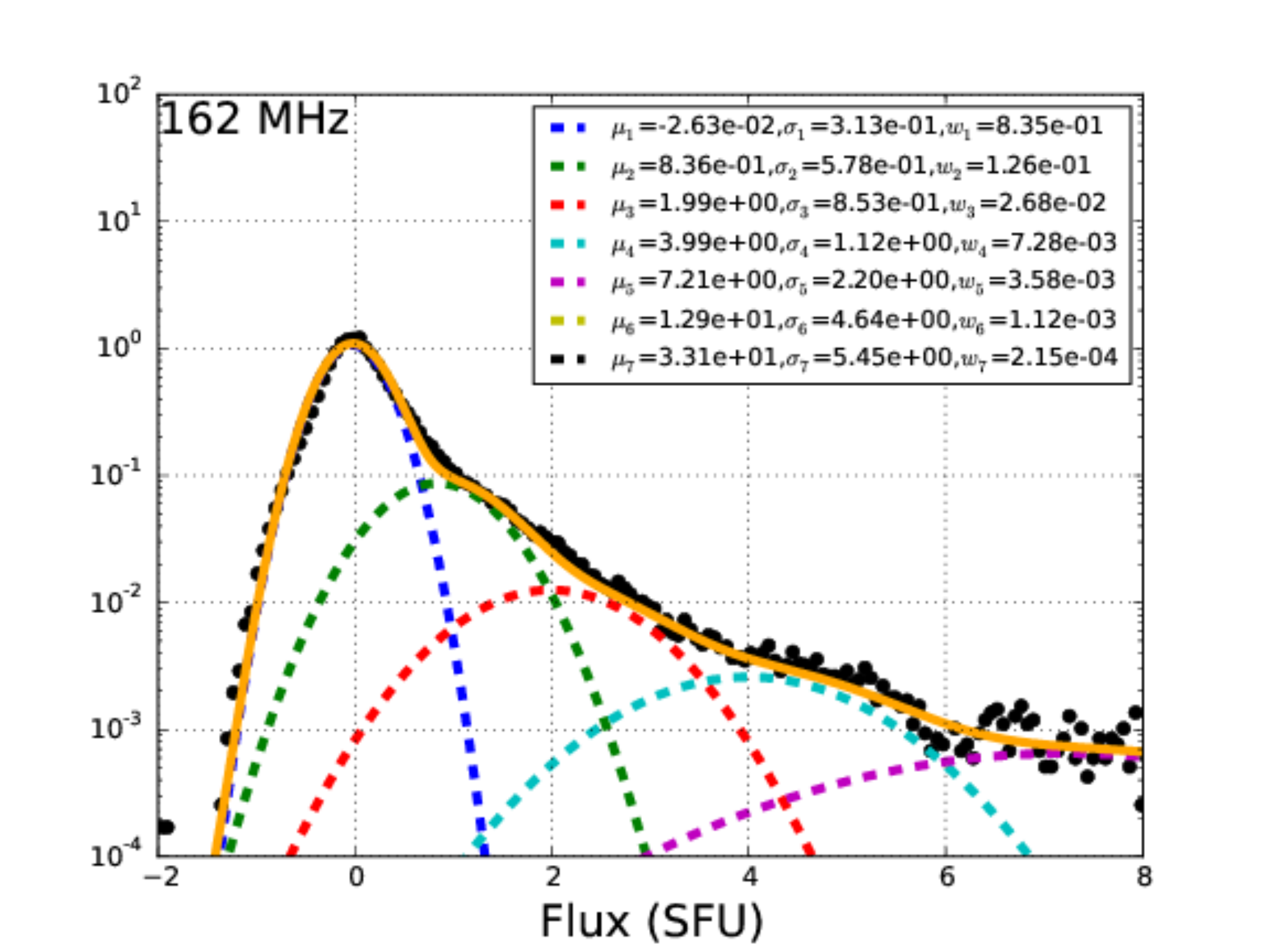}
\caption{Histogram for baseline Tile011-Tile023 for one hour of data at $\sim$162 MHz.
The black circles show the observed distribution of solar flux density in SFU.
The dashed curves show the individual component Gaussians and the solid orange curve, the sum of all the model components.
The y axis is in log scale and the observed distribution spans four orders of magnitude.
The best-fit parameters of all Gaussian components are also listed.
\label{Fig:hist}}
\end{figure}

\begin{figure}%[htbp]
\begin{center}
\begin{tabular}{cc}
\includegraphics[scale=0.35]{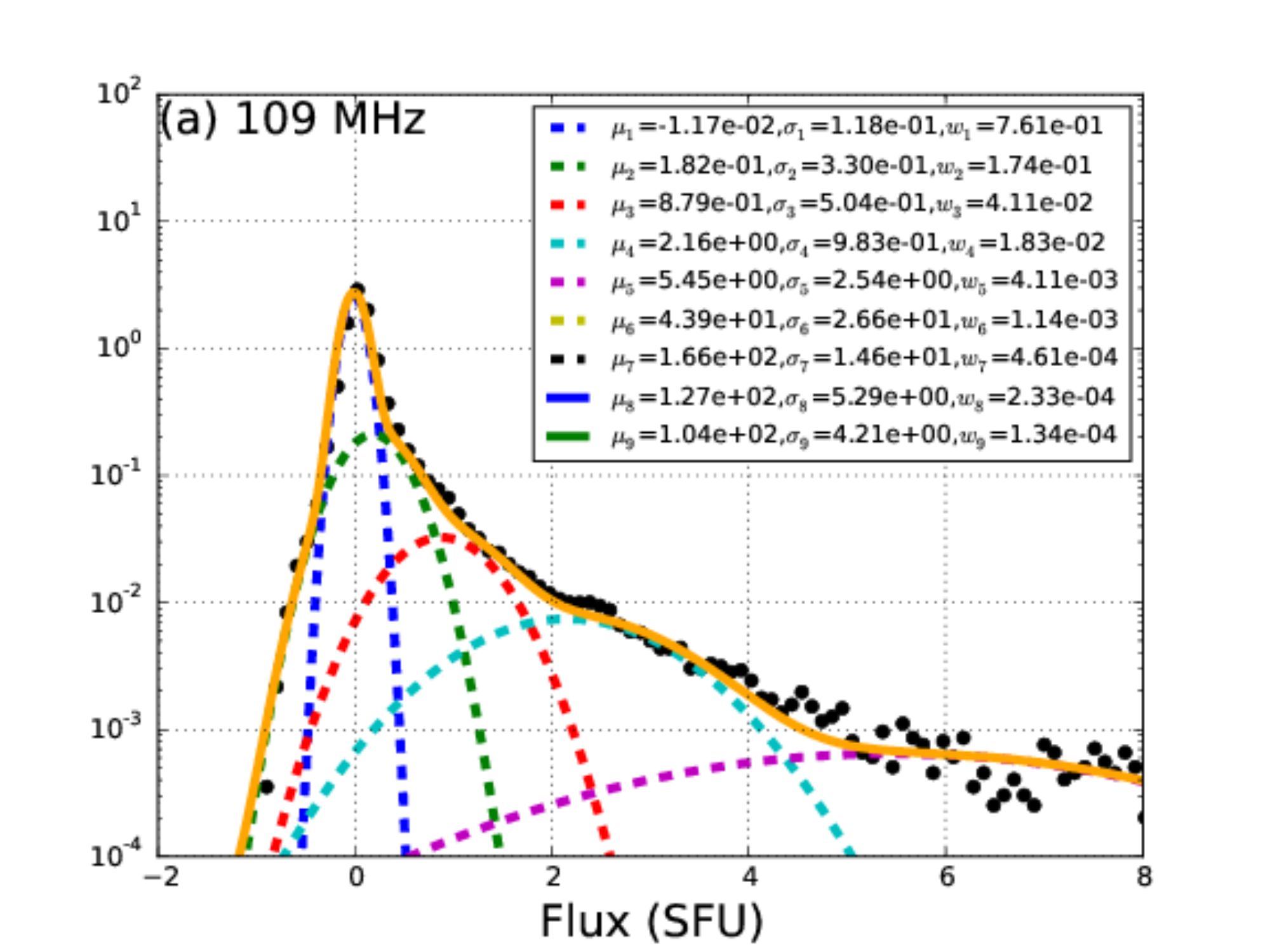} &
\includegraphics[scale=0.35]{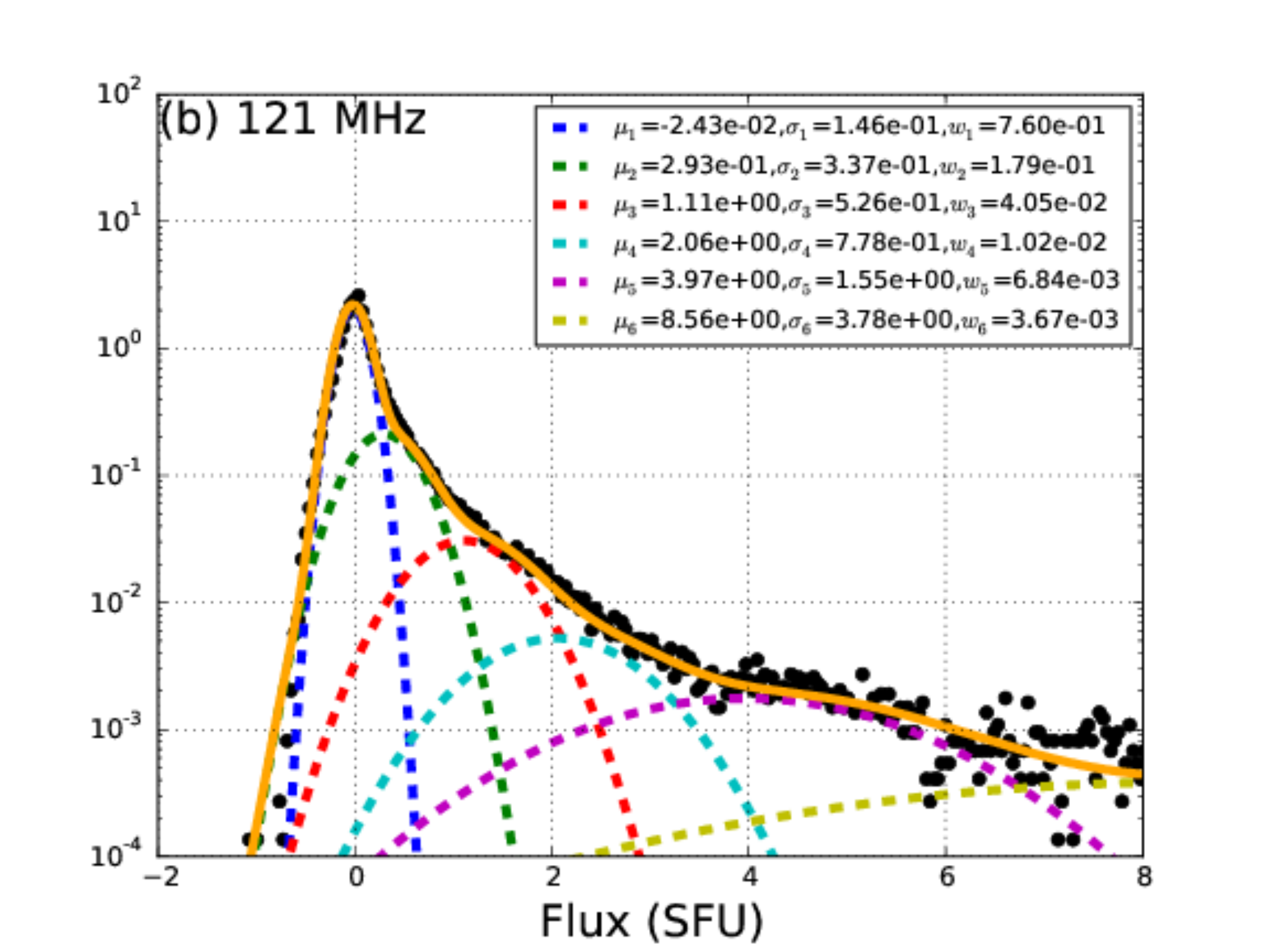} \\
%(a) 109 MHz&(b) 121 MHz \\
\includegraphics[scale=0.35]{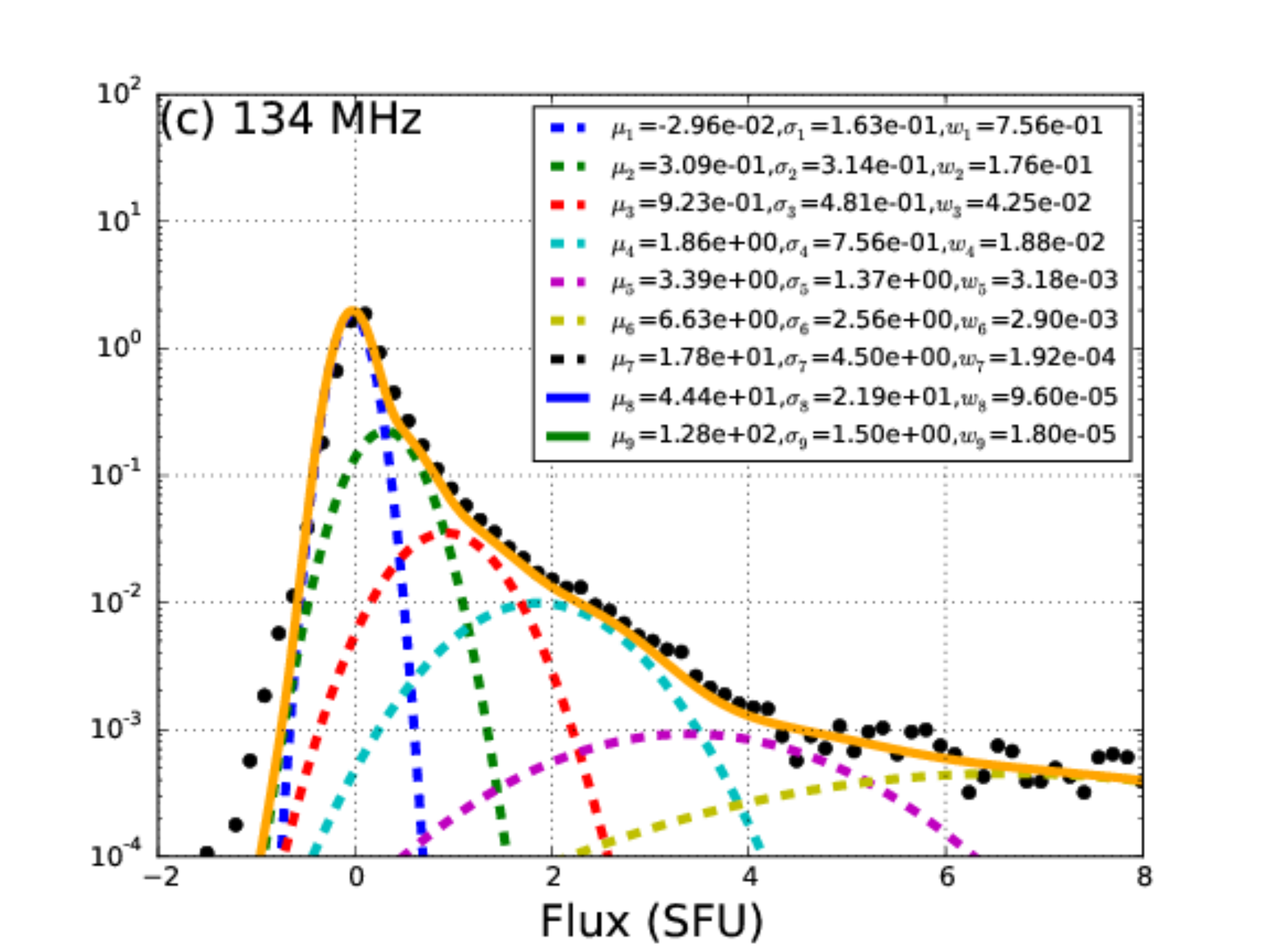}  &
%(a) 109 MHz&(b) 121 MHz & (c) 134 MHz\\
\includegraphics[scale=0.35]{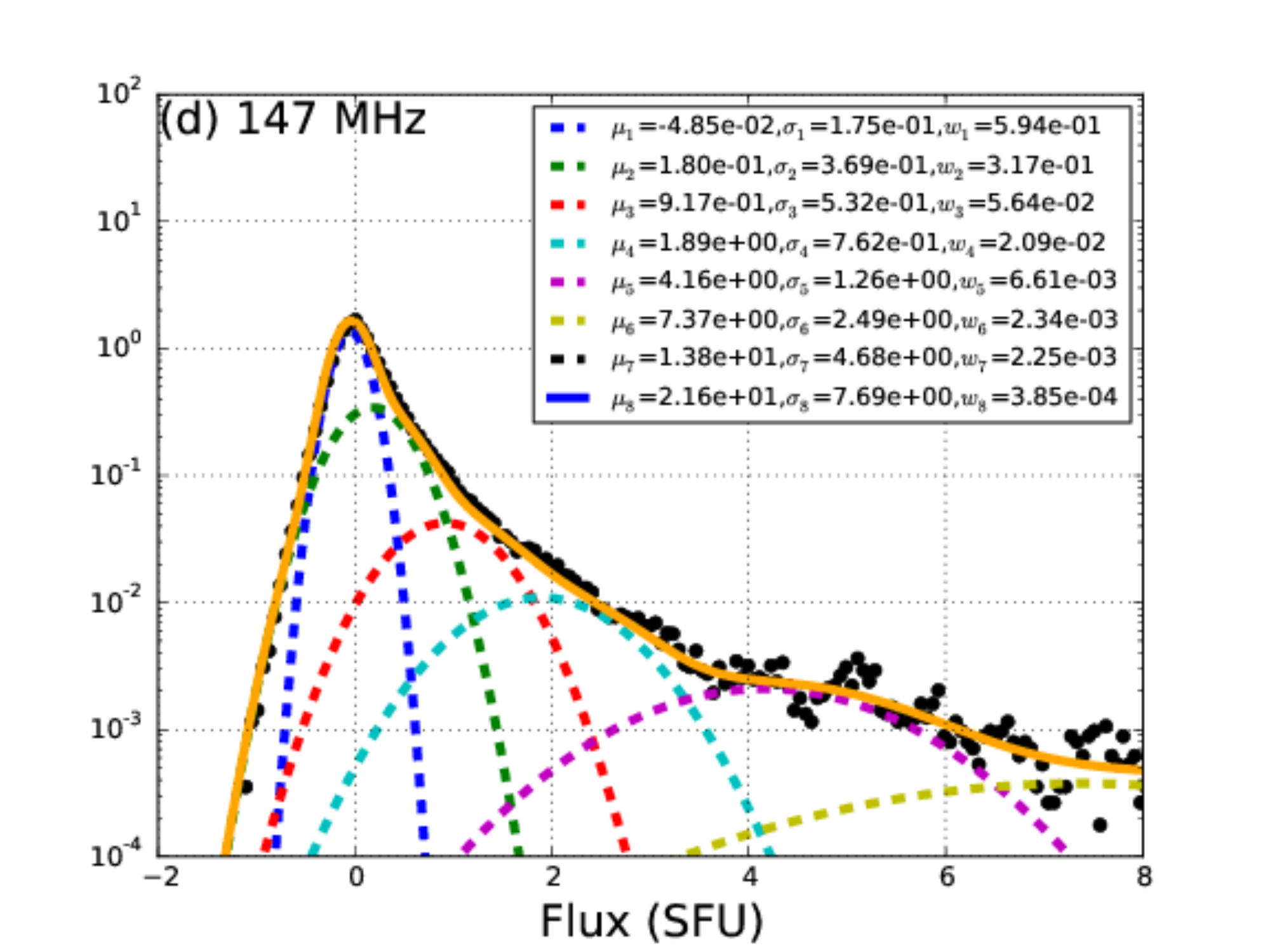} \\
%  (a) 134 MHz&(b) 147 MHz \\
\includegraphics[scale=0.35]{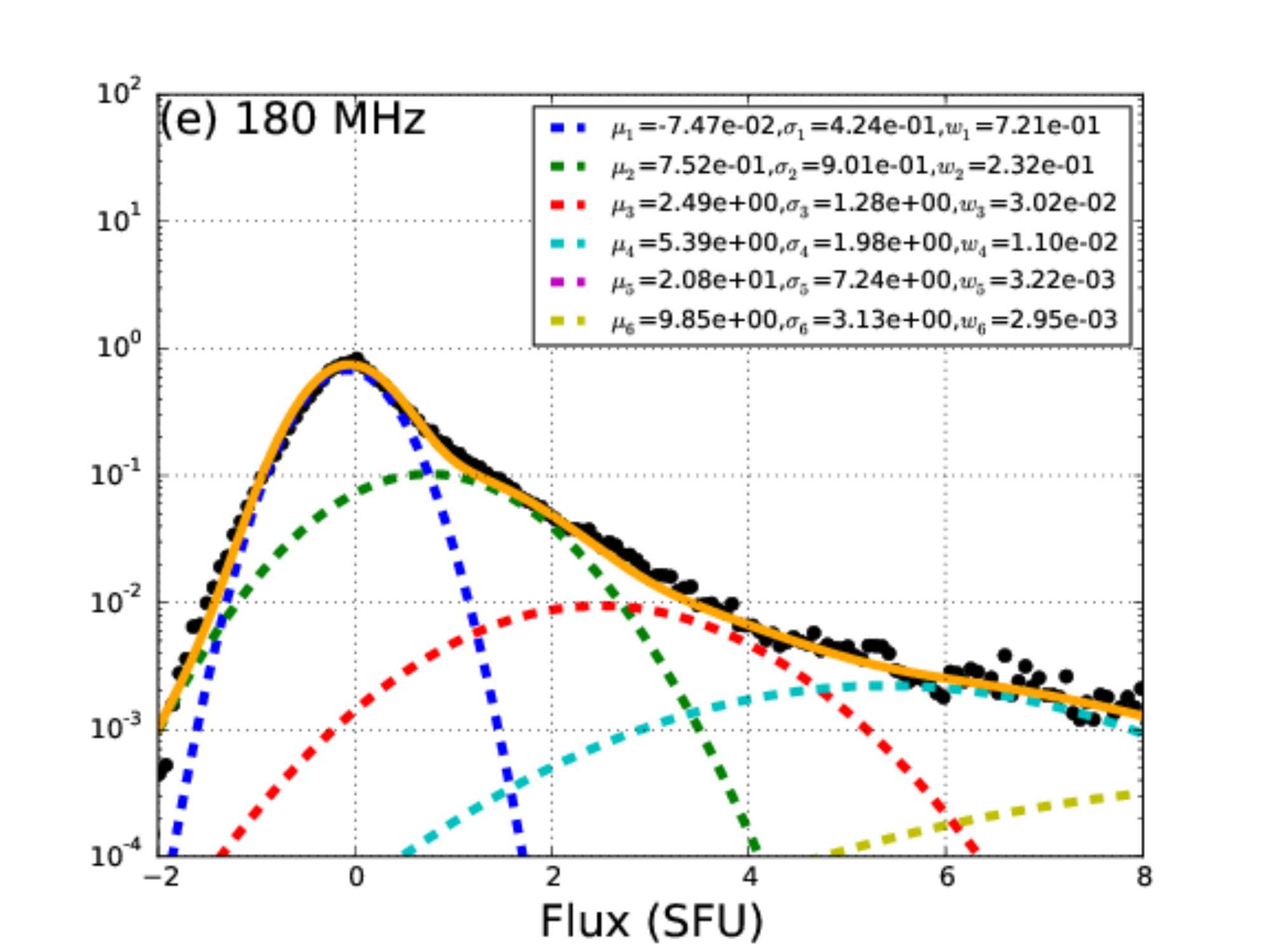} &
%(d) 147 MHz&(e) 162 MHz & (f) 180 MHz\\
\includegraphics[scale=0.35]{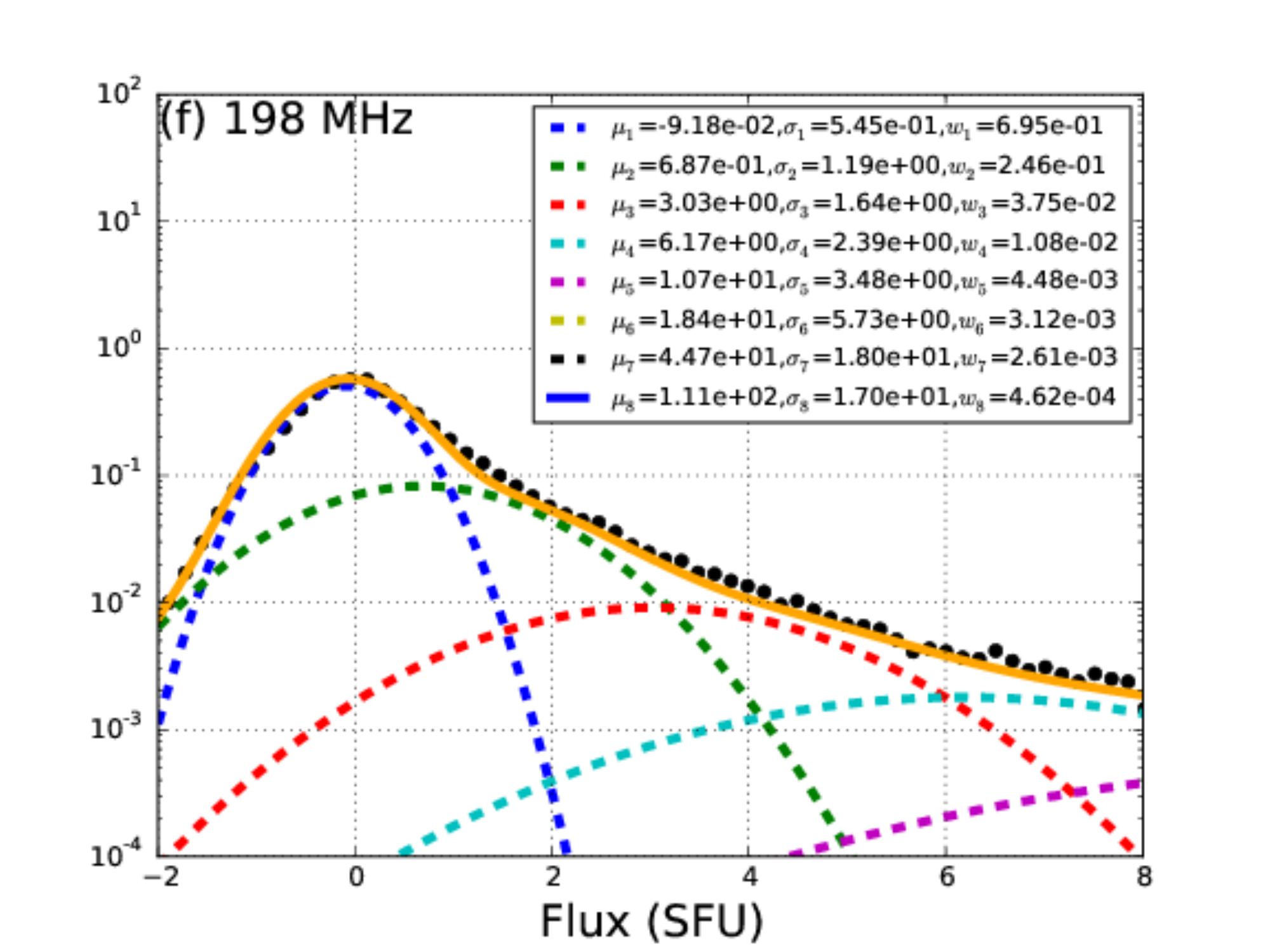} \\
%  (a) 180 MHz&(b) 198 MHz \\
\includegraphics[scale=0.35]{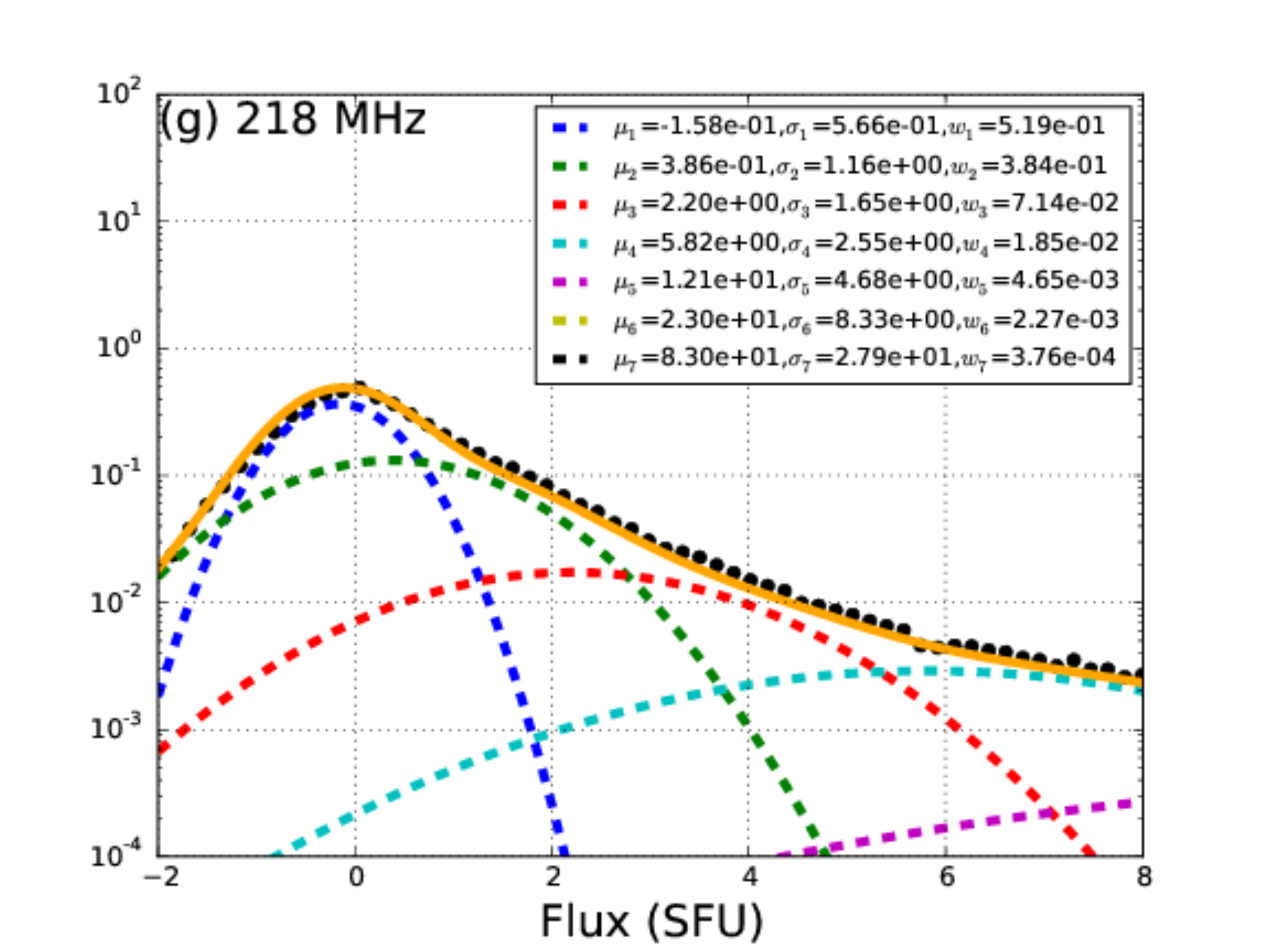} &

\includegraphics[scale=0.35]{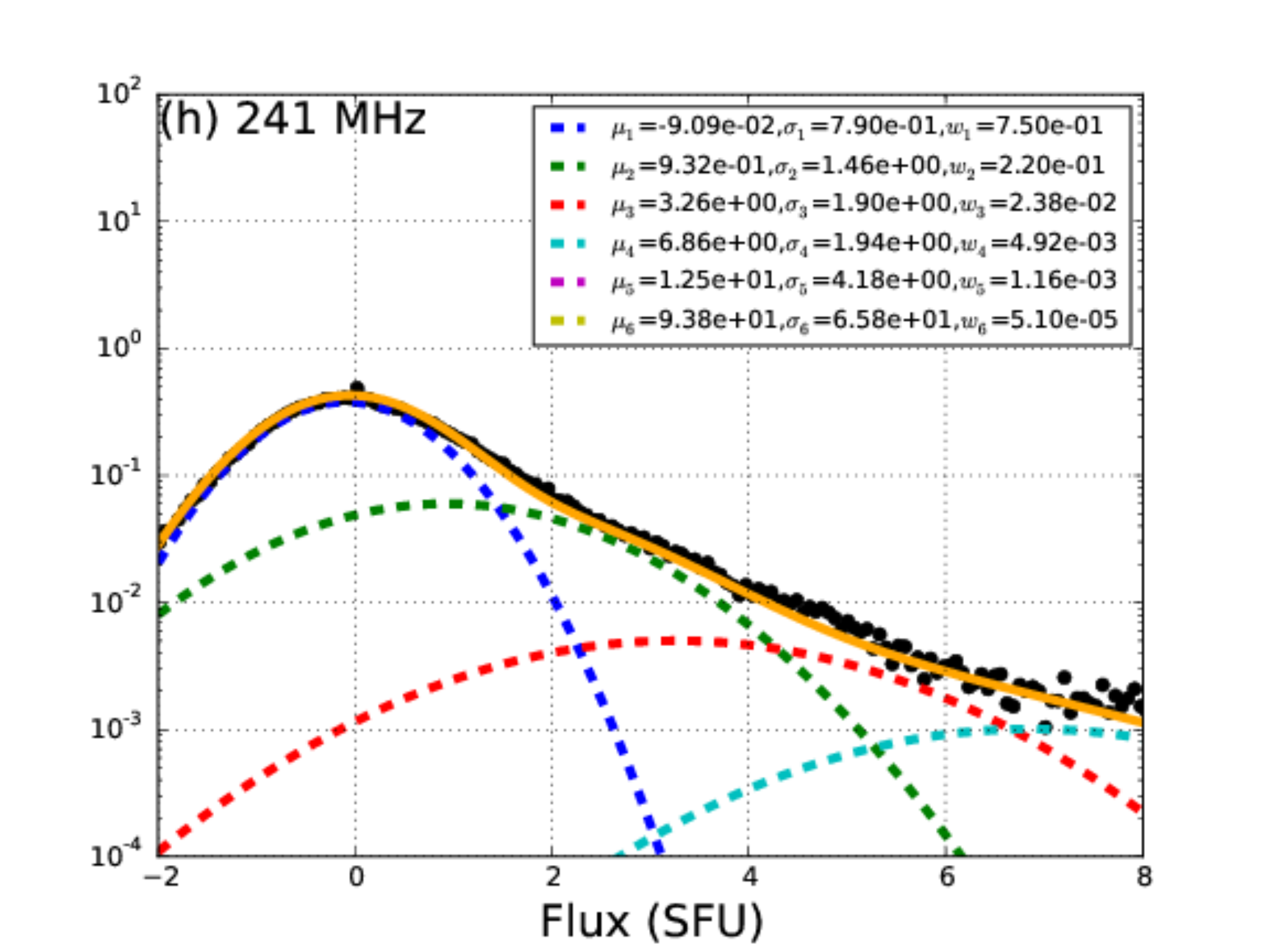} \\
%(h) 218 MHz & (i) 241 MHz\\
\end{tabular}
\end{center}
\caption{The flux density histograms for Tile011-Til023 baseline for other frequencies in the same format as Fig.\ \ref{Fig:hist}.
\label{Fig:hists}}
\end{figure}

For each of the nine spectral bands, 28 spectral channels were grouped together for Gaussian mixture modeling.
These channels were chosen to exclude the edges of the coarse spectral channels band which are affected by instrumental artefacts.
This analysis was carried out for all six independent baselines for each of the spectral bands to provide a check for consistency.
We first describe the results obtained for the band centered $\sim$162 MHz for one baseline in detail.
Similar results were obtained following an identical procedure for the other spectral bands and baselines as well.
The BIC criterion indicates that the best-fit model for these data has seven Gaussian components.
The data and the best-fit Gaussian model are shown in Fig.\ \ref{Fig:hist} for the baseline Tile011-Tile023 along with the values of $\mu$s, $\sigma$s and $w$s for each of the component Gaussians. 
The EM process showed a smooth and robust convergence to stable values and, in this instance, took 320 iterations to reach the convergence criterion (for a Gaussian parameter $p$ at $j^{th}$ iteration, $p_{j} - p_{j-1} <$ 10$^{-6}$). 
The Gaussian component with the largest weight ($0.835$) has mean $\mu$ close to zero ($-0.026$ SFU), consistent with our expectations.
For these data, the Gaussian component with the largest $w$ is always found to have the smallest $\mu$ as well.
We label the parameters of the Gaussian components with subscripts in descending order of their $w$s and interpret the component with index $i=1$ as the one corresponding to the slowly varying thermal emission.
The weights $w$ of the component Gaussians decrease monotonically with increasing $\mu$, and this trend continues even after the $w$s fall to a fraction of a percent of the data.
%%----------------------------------------------------------------------
%%----------------------------------------------------------------------
The fitted GMMs for all other frequencies for the same baseline are shown in Fig.\ \ref{Fig:hists}; these follow trends similar to those seen in Fig.\ \ref{Fig:hist}.

\begin{figure}%[htbp]
\begin{center}
\begin{tabular}{ccc}
\includegraphics[scale=0.25]{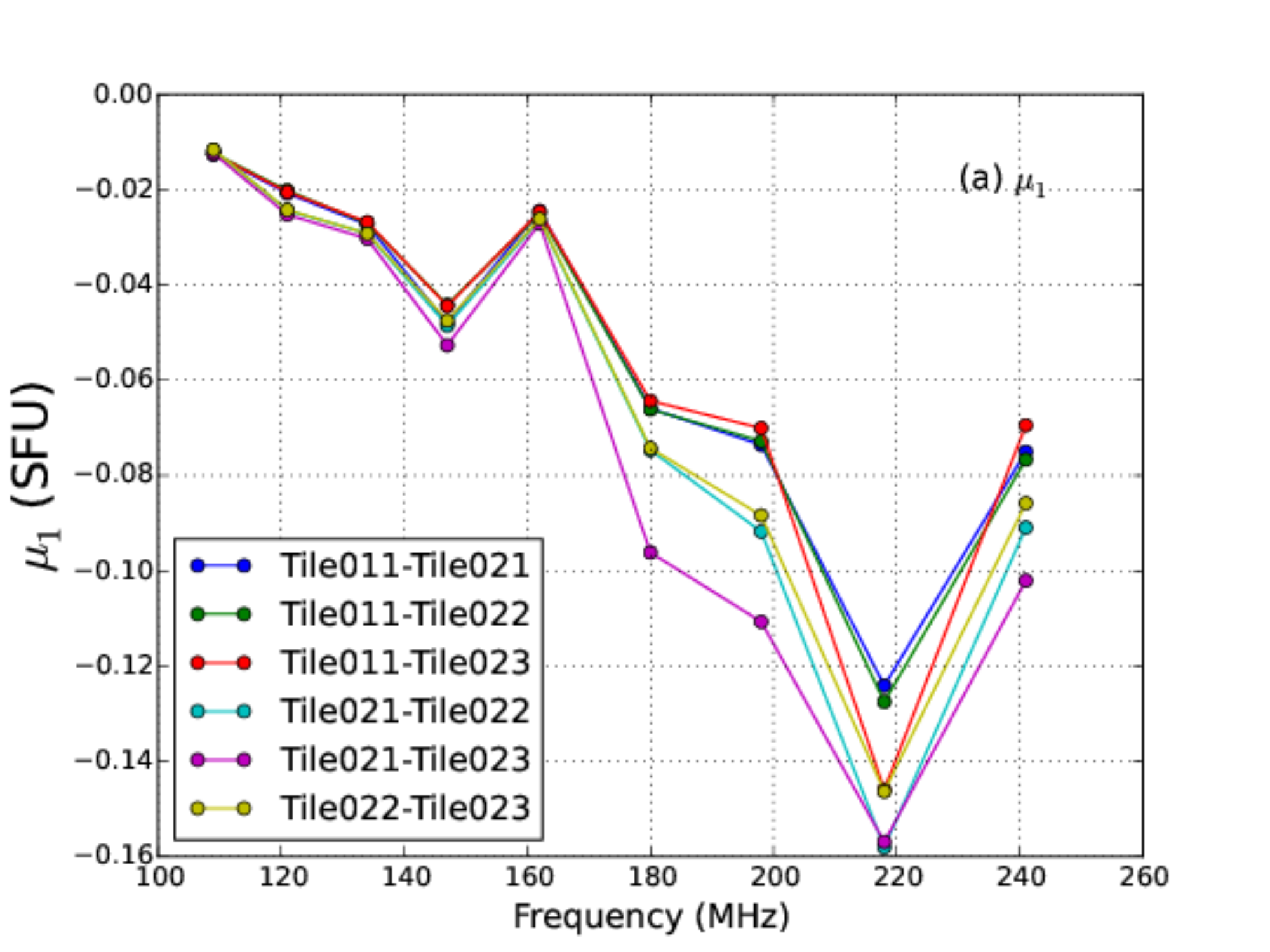}
 &
\includegraphics[scale=0.2]{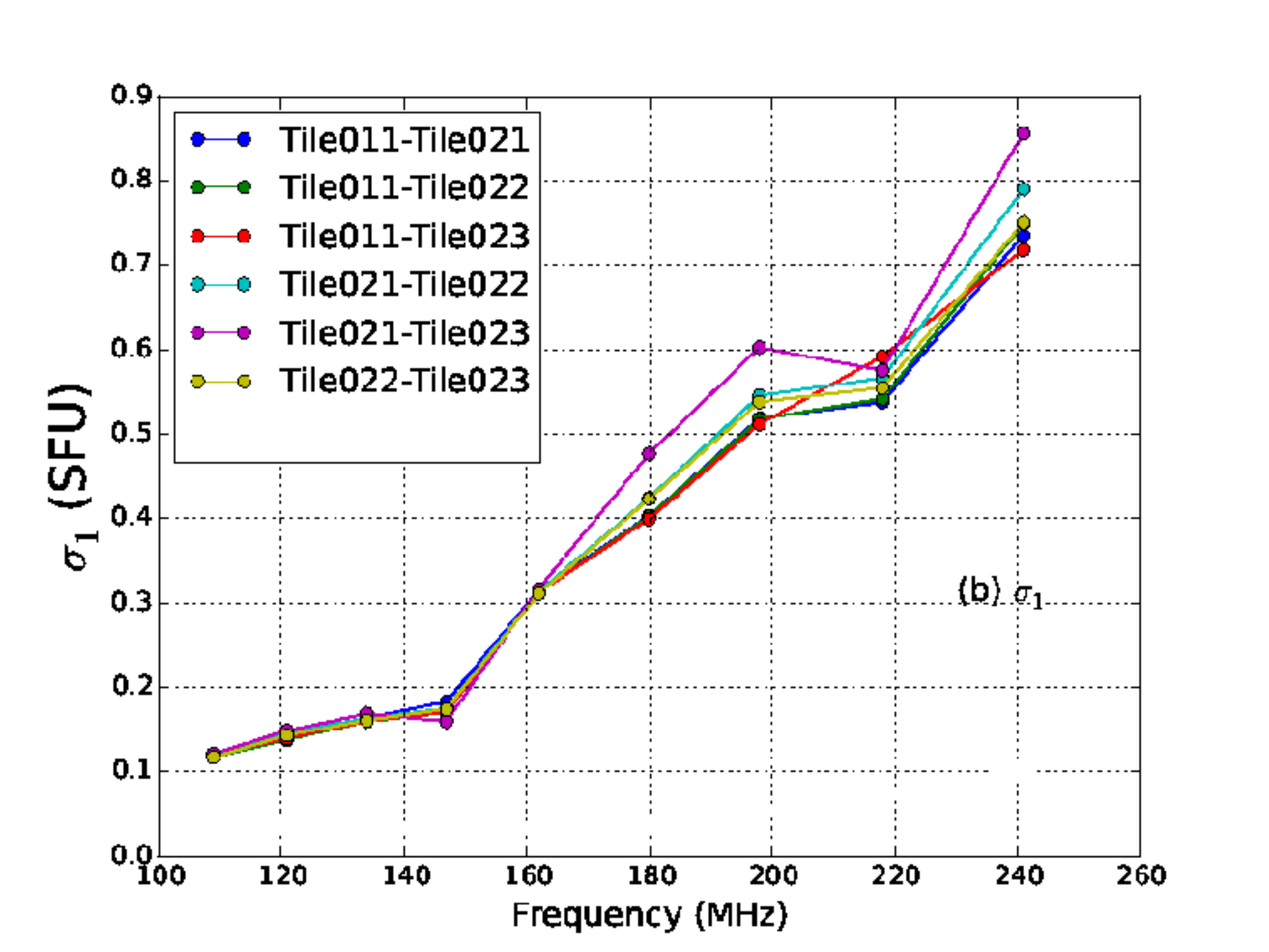}
 &
\includegraphics[scale=0.25]{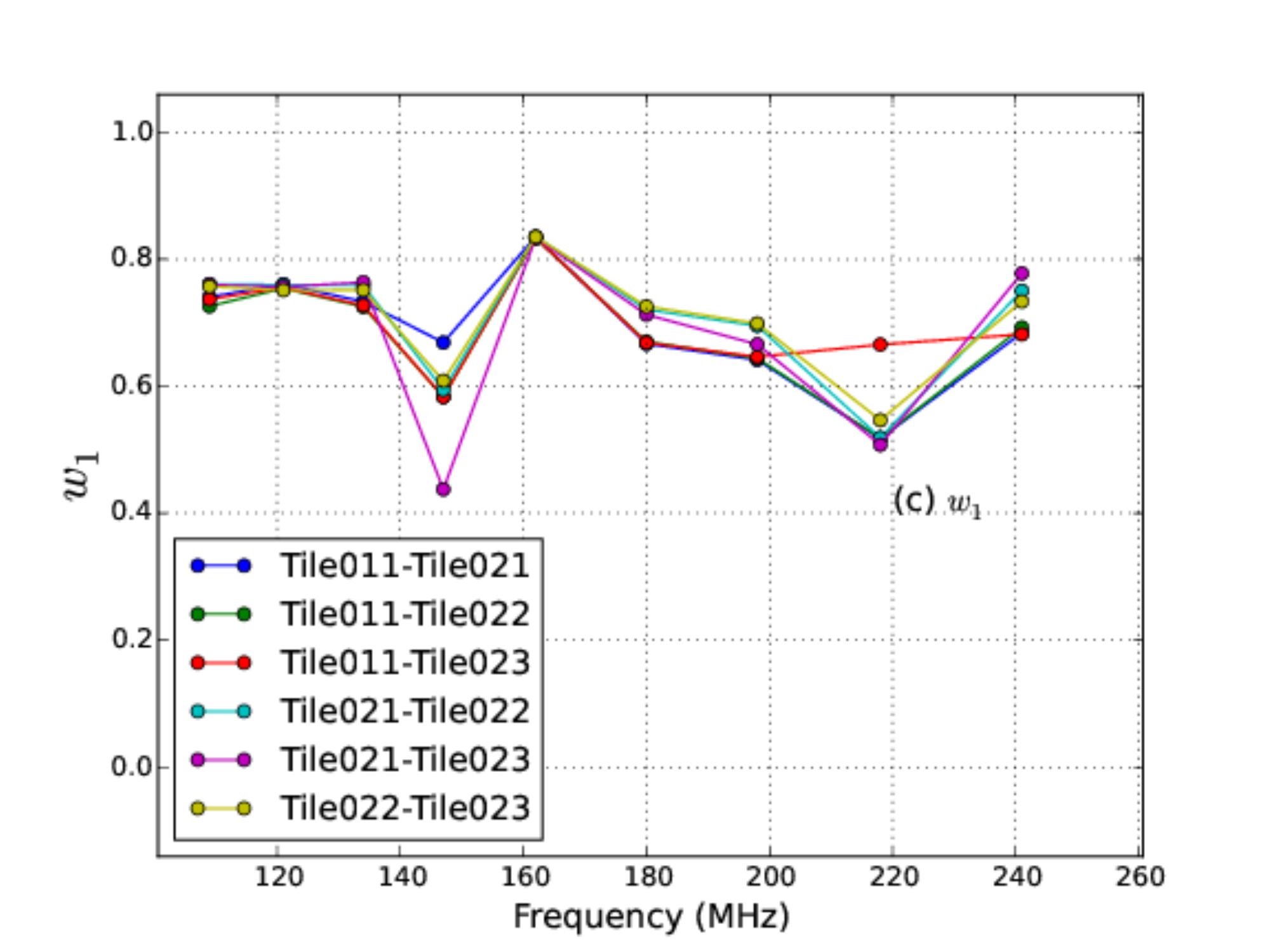} \\
\end{tabular}
\end{center}
\caption{From left to right, the panels show the GMM estimates for $\mu_1$, $\sigma_1$ and $w_1$ as a function of frequency for all the six baselines.
The middle panel also shows the expected thermal noise ($\delta S_{th}$) as a function of frequency in black.   
\label{Fig:result_pguass}}
\end{figure}

Figure \ref{Fig:result_pguass} shows the variation of $\mu_{1}$, $\sigma_{1}$ and $w_{1}$ with frequency, for all the six baselines considered here. %listed in Table. \ref{Tab:baselines}. 
Though, $\mu_{1}$s show a weak trend with frequency consistent across baselines, they are all close to zero with values usually within 1\% of the mean of the non-impulsive flux density (Table \ref{Tab:non-thermal}). 
This confirms that the median filtering technique used here has been effective in estimating and removing the slowly varying thermal flux density.
The $\sigma_{1}$s increase monotonically with frequency. % with low scatter between the values estimated from different baselines.
%The middle panel of Fig.\ \ref{Fig:result_pguass} also shows the theoretical thermal noise, $\sigma_{Th}$, estimated based on the expected values of system temperature, T$_{Sys}$, and effective area, $A_{eff}$, and is given by the following equation {\bf Cite Synthesis Imaging text book}: 
%\begin{equation}
%\sigma_{Th} = \frac{2\ k}{A_{eff}} \frac{T_{Sys}}{\sqrt{\Delta \nu\ \Delta t}},
%\label{Eq:Thermal_uncertainty}
%\end{equation}
%where $\Delta \nu$ and $\Delta t$ represent the frequency and time integration of individual visibilities.
%$\sigma_{Th}$ varies from 0.06 SFU at 109 MHz to 0.16 SFU at 241 MHz with a shallow minima at $\sim$160 MHz. 
%The observed $\sigma_1$ is always significantly larger than $\sigma_{Th}$. 
%implying that $\sigma_1$ is not determined solely by $\sigma_{Th}$ and there must be other processes making considerably larger contributions to it. 
The mean value of $w_1$s varies between $\sim$0.5 -- 0.8 at different frequency bands and does not seem to follow any specific trend with frequency. 
Data from different baselines constitute independent realizations drawn from the same distribution.
The estimates of all the three parameters are remarkably consistent between baselines, demonstrating the robustness of GMM-based methodology followed here.

%%------------------------------------------------------

\section{Uncertainty estimates}
\label{Sec:uncert-estimates}

It is evident from Figs.\ \ref{Fig:hist} and \ref{Fig:hists} that a superposition of Gaussian components model the observed $S_{\odot}$ distribution remarkably well, and  Fig.\ \ref{Fig:result_pguass} demonstrates that different baselines with independent datasets also lead to remarkably consistent estimates for parameters of the first Gaussian.
As in all problems of this nature, it is important to establish uniqueness of the best-fit Gaussian parameters determined by GMM.
In this section we describe our efforts to check for degeneracy in the best-fit Gaussian parameters and also build a quantitative estimate the level of uncertainty associated with them.
We focus on uncertainty in parameters of the first Gaussian as they are the only ones of physical interest in our context.

\subsection{Varying initial guesses}
\label{Subsec:varying-initial-guesses}
\begin{figure}%[htbp]
\begin{center}
\includegraphics[scale=0.25]{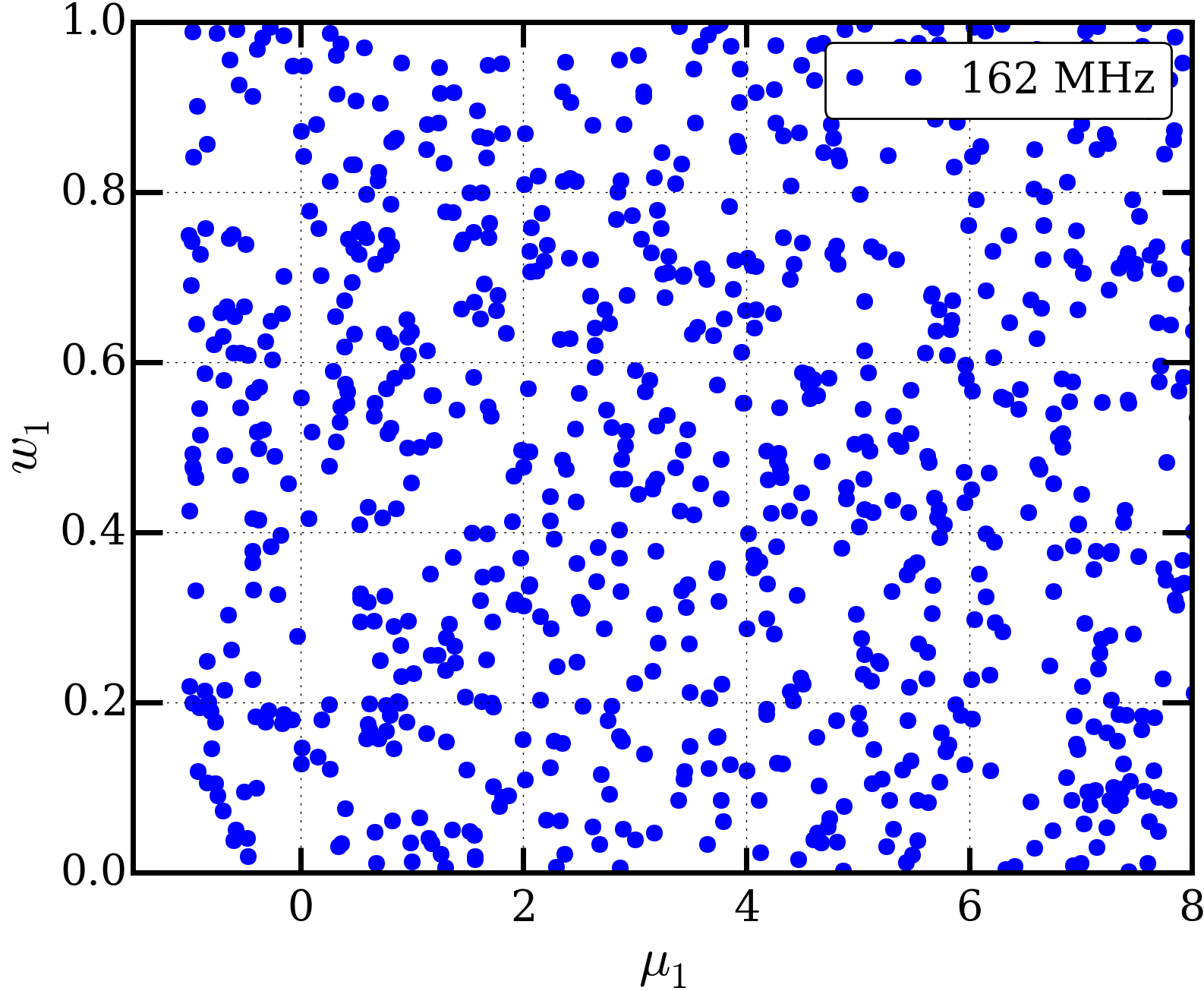} 
\includegraphics[scale=0.25]{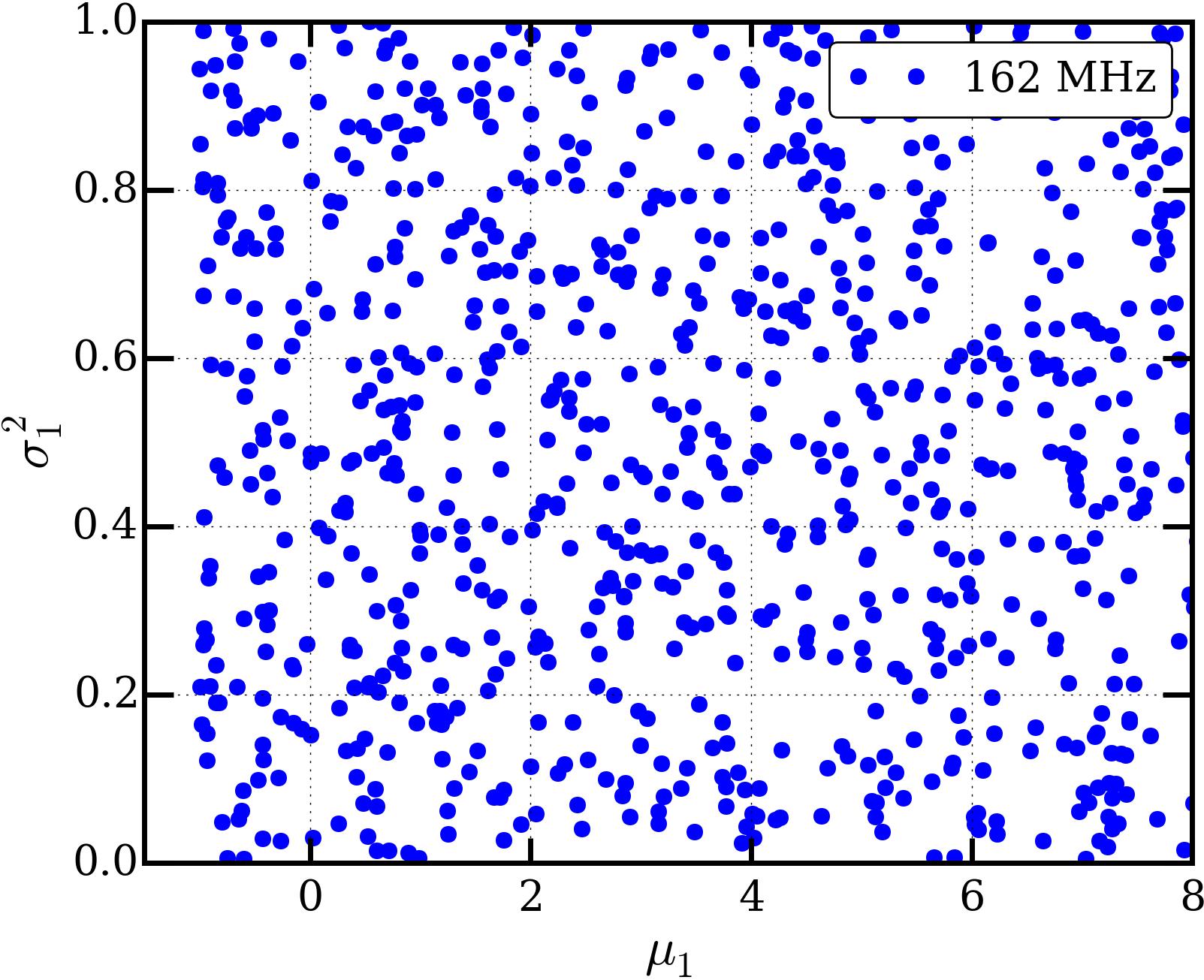}
\end{center}
\caption{
The left and right panels show the distribution of initial guesses used to seed the GMM in the $\mu_1$--$w_1$ and $\mu_1$--$\sigma_1^2$ planes, respectively for 162 MHz and baseline Tile011-Tile022.
\label{Fig:initial-guess}}
\end{figure}

We first check for the dependence of the final GMM parameters on the choice of the initial guesses for these parameters supplied to the EM algorithm.
Sensitive dependence on the initial guess can often be the case when the number of degrees-of-freedom (DoF) available to the algorithm are large and not well constrained, leading to an over-fitting of the data.
Though the use of BIC is intended precisely to limit the number of available DoF and avoid such scenarios, it is useful to explicitly verify this.
If there are indeed multiple local minima in the minus-log-likelihood hypersurface, seeding the algorithm with different initial guesses spanning a large range can drive the algorithm to different minima. 
This can lead to a dependence of the final best-fit parameters on the choice of the initial guesses.
To test for this possibility, we ran GMM on the same dataset using 1000 different sets of initial guesses to seed the EM algorithm and allowed it converge to its final values, i.e for a Gaussian parameter $p$ at $j^{th}$ iteration, $p_{j} - p_{j-1} <$ 10$^{-6}$. 
The boundaries of the parameter space for the initial guesses were determined by the range of the data and the guesses were drawn from a uniform distribution in this space \citep{Karlis2003}.
An example of the distribution of initial guesses for a dataset corresponding to one of the frequencies is shown in Fig.\ \ref{Fig:initial-guess}.

\begin{figure}%[htbp]
\begin{center}
\begin{tabular}{ccc}
\includegraphics[scale=0.25]{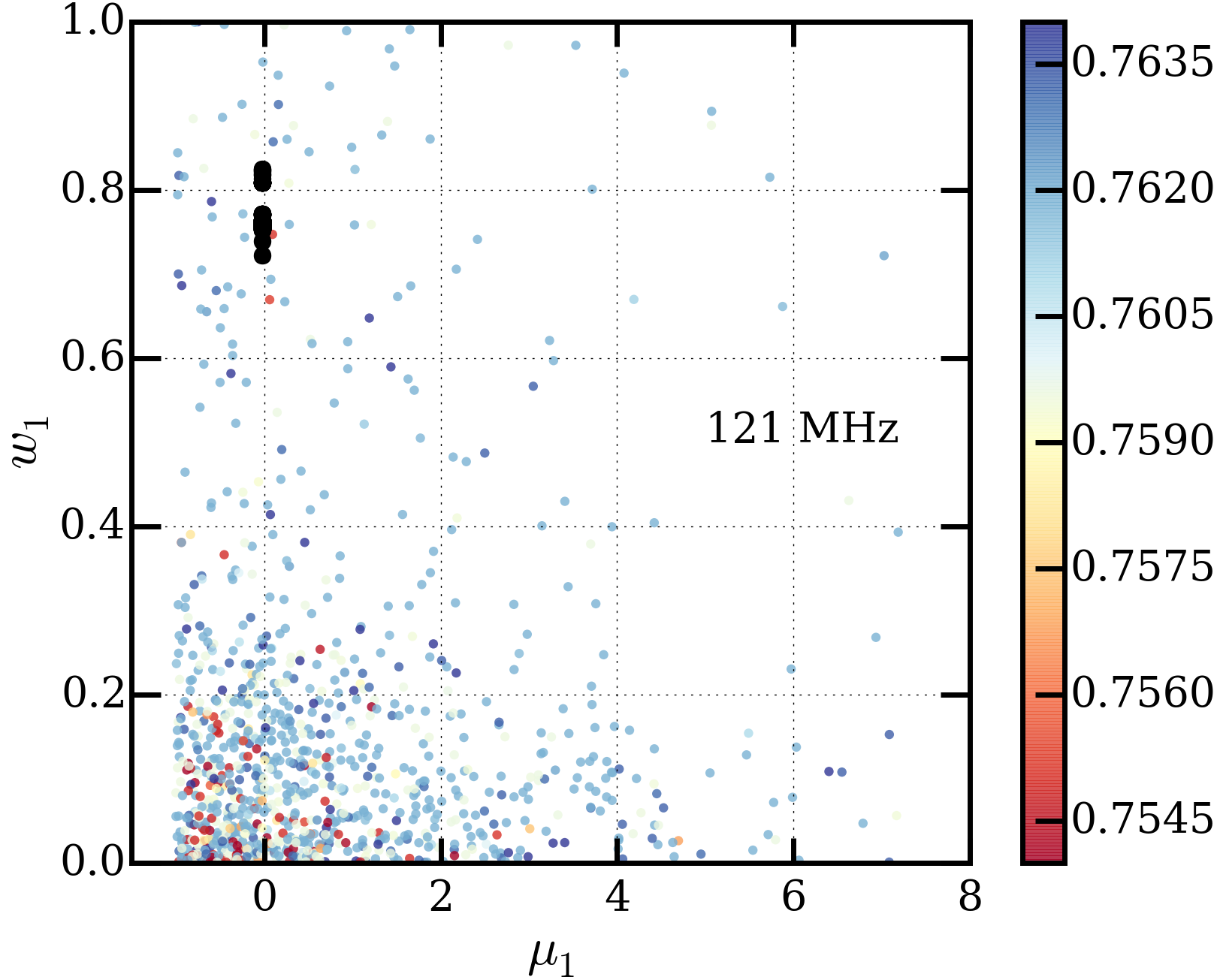} 
 &
\includegraphics[scale=0.25]{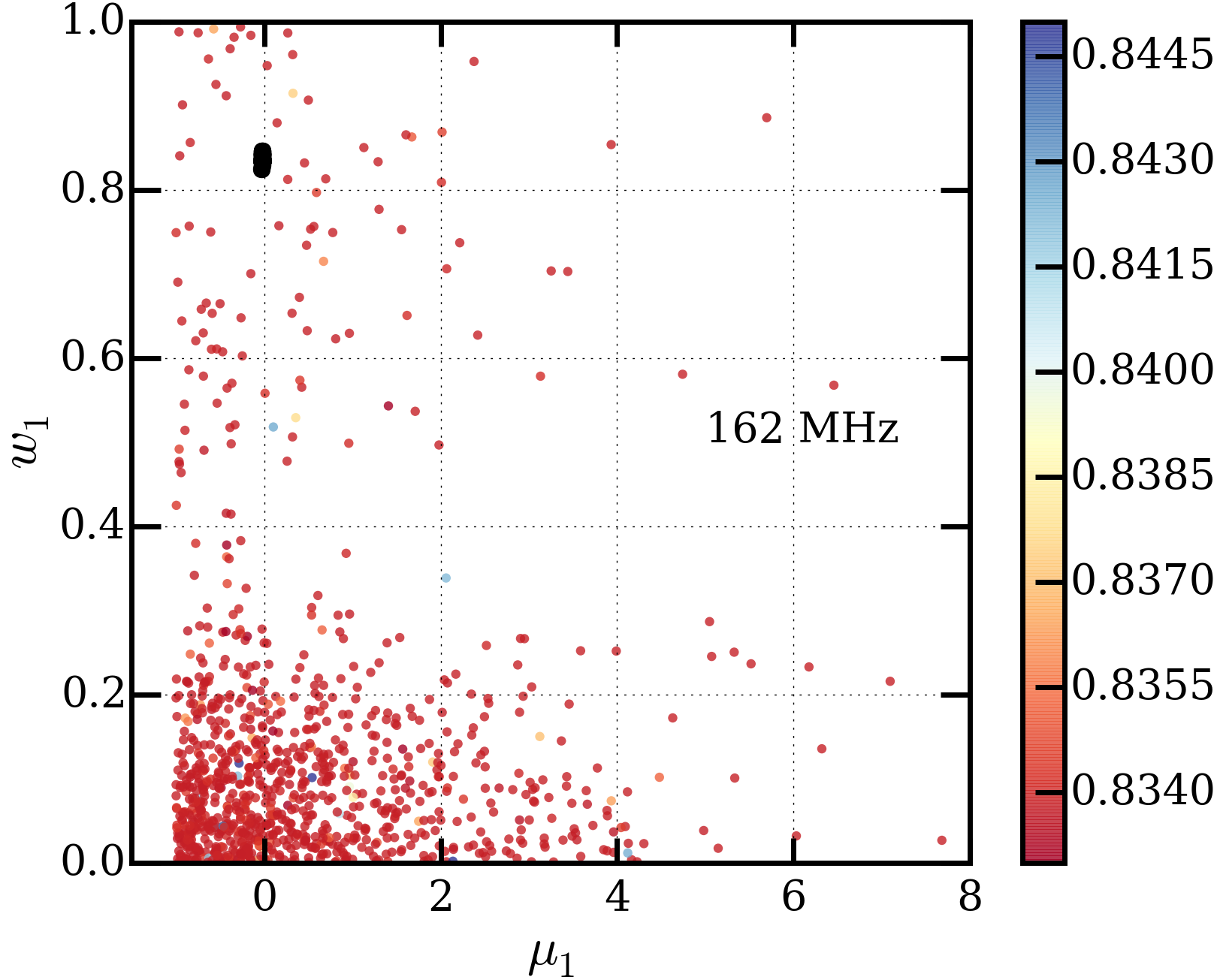}
 &
\includegraphics[scale=0.25]{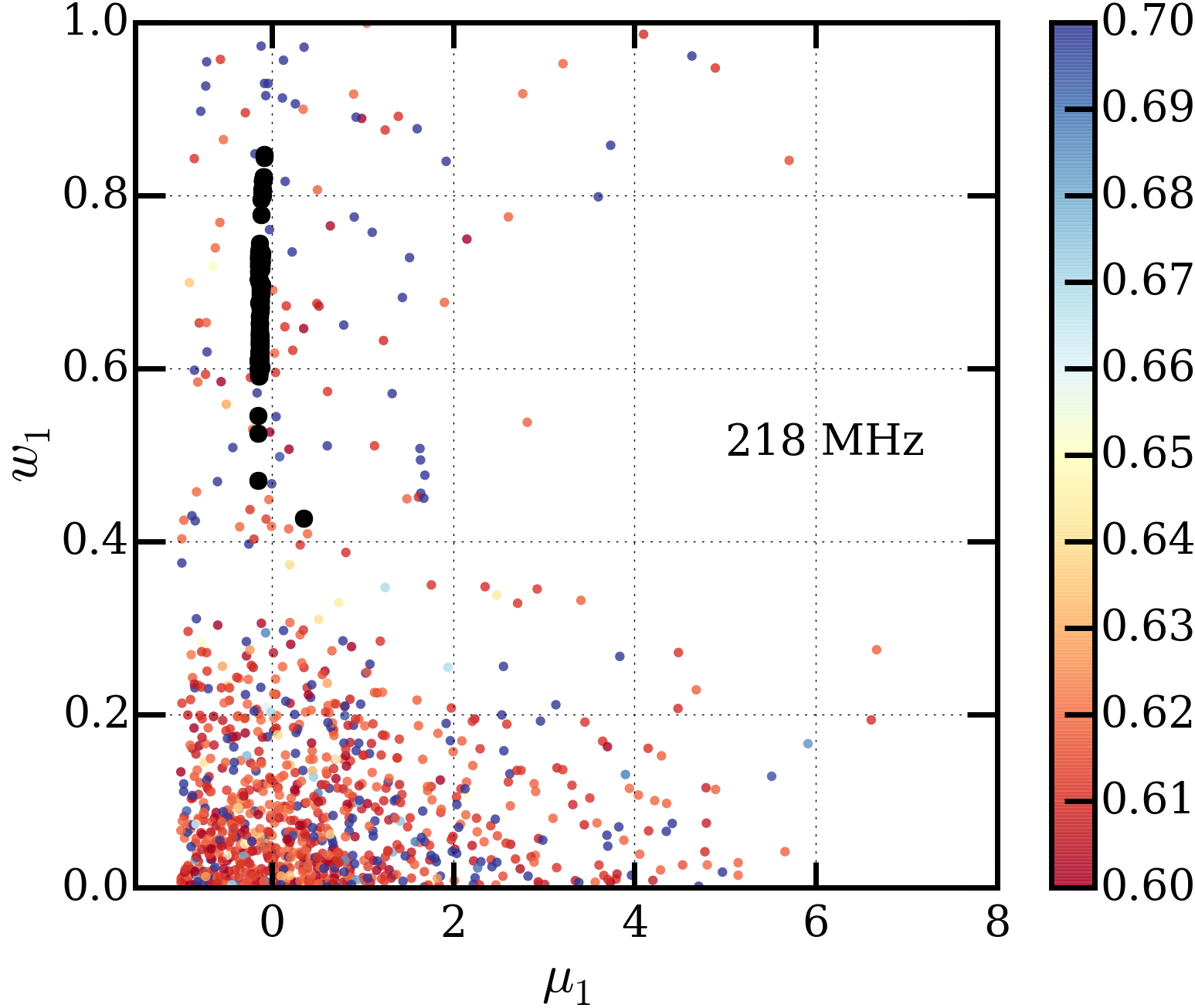}\\

\includegraphics[scale=0.25]{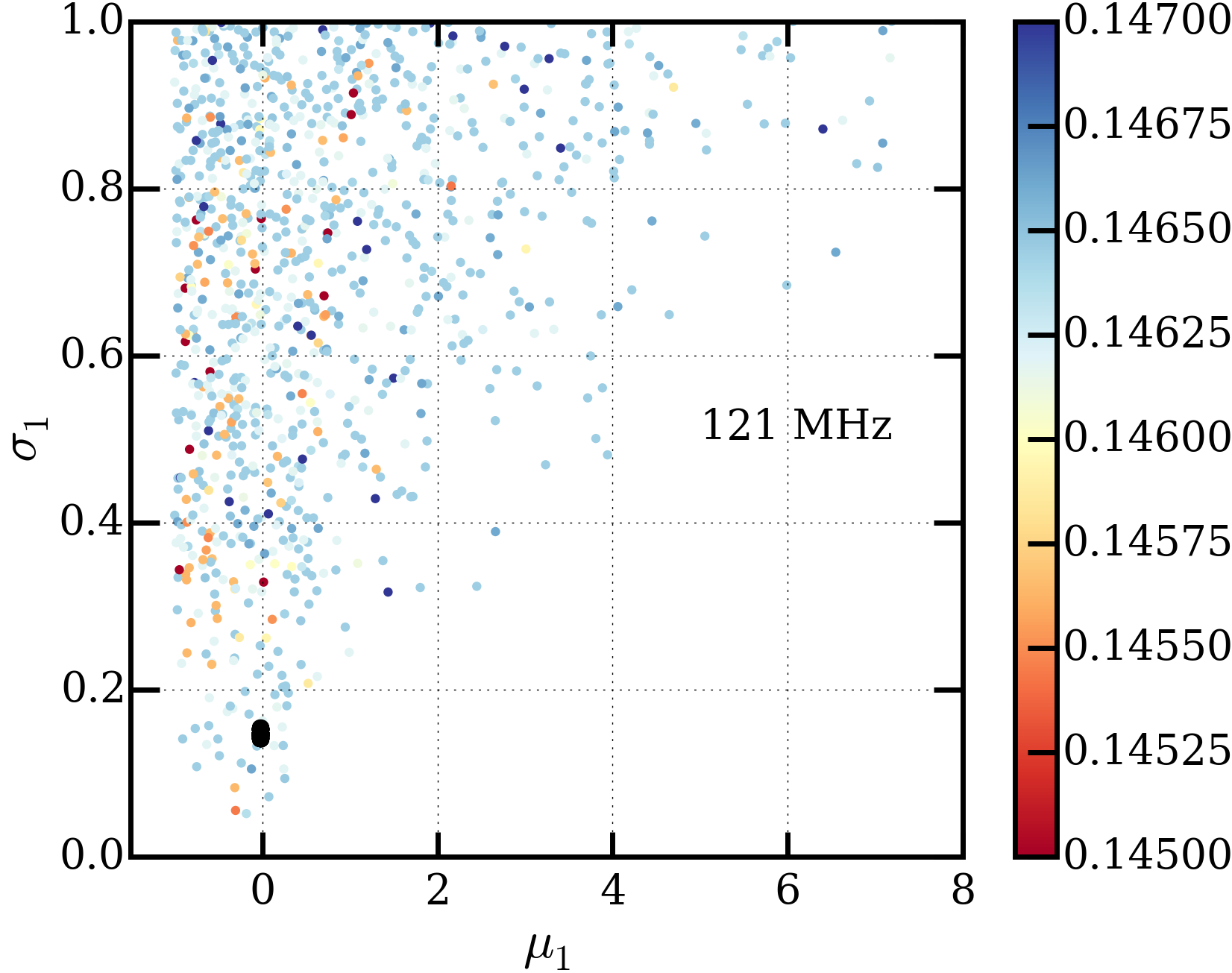}
 &
\includegraphics[scale=0.25]{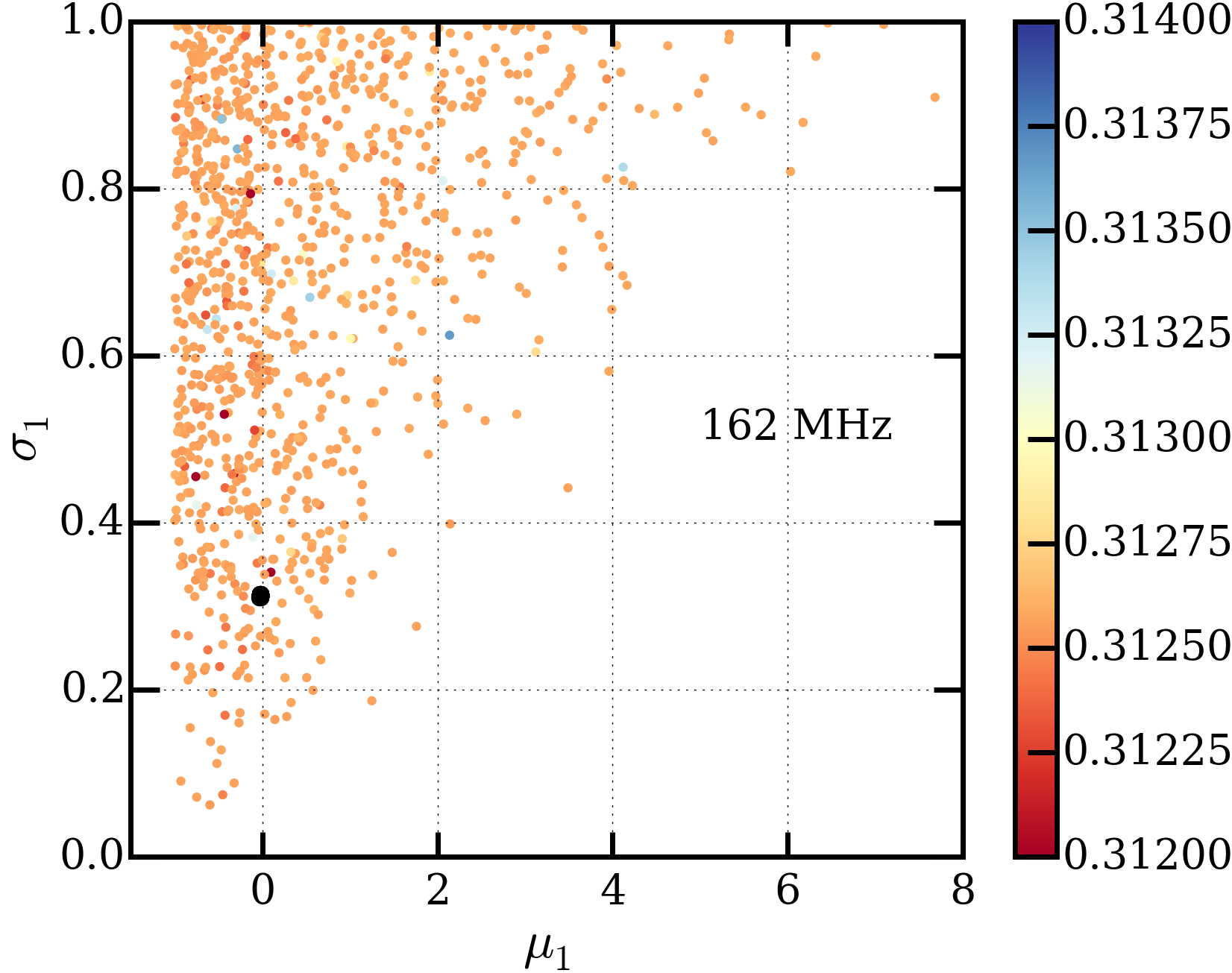}
 &
\includegraphics[scale=0.25]{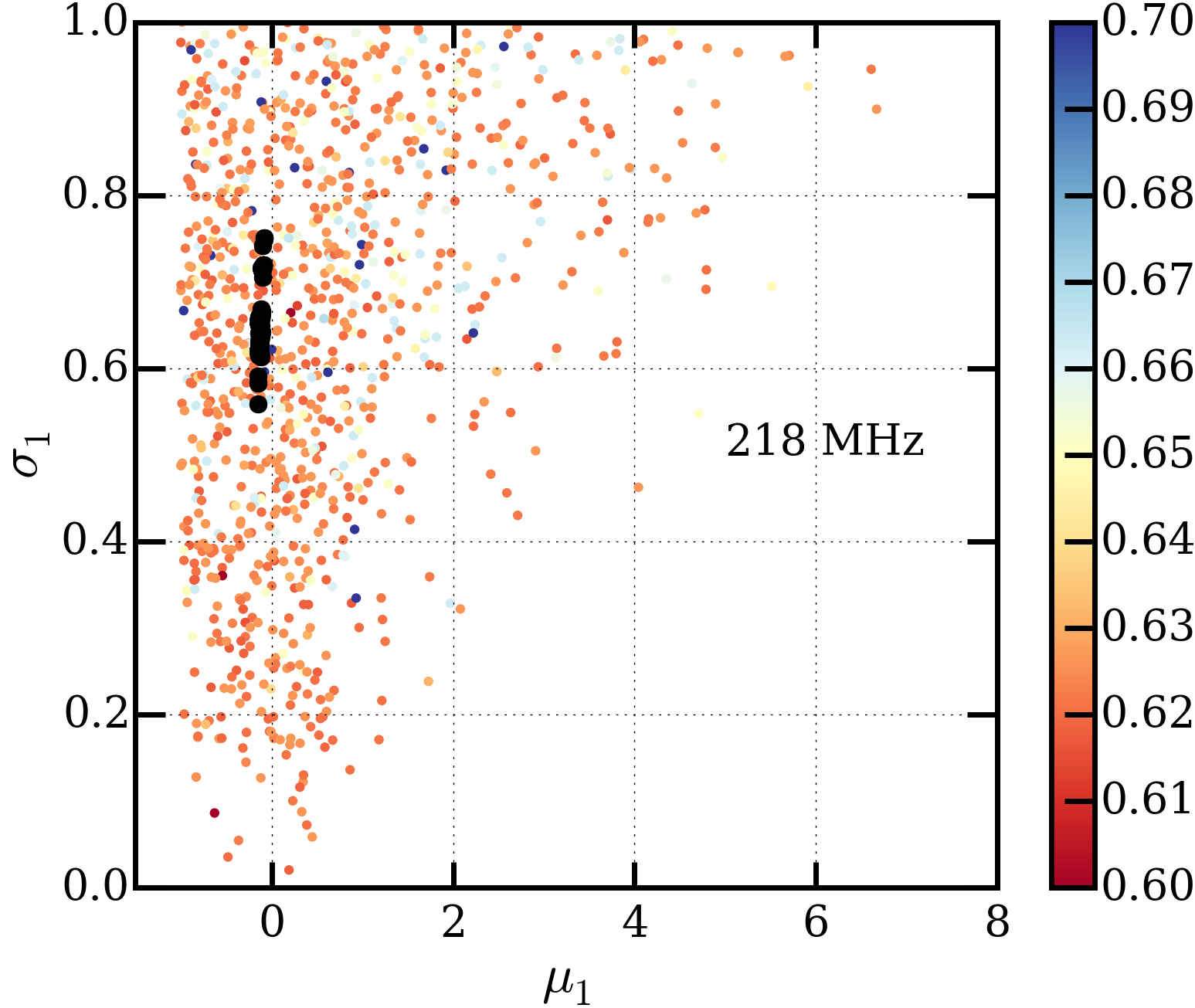}
\end{tabular}
\end{center}
\caption{
The colored symbols show the value of the initial guesses provided to the GMM and the black circles, the final values to which it converged.
The top row shows the $\mu_1$--$w_1$ plane and the bottom row the $\mu_1$--$\sigma_1$ plane.
From left to right, the columns correspond to observing frequencies centered at 121, 162 and 218 MHz respectively.
The colors of the symbols indicate the final values of $w_1$ to which that particular initial guess converged.
The colorbar has been auto-scaled to cover the entire range of value spanned by the final value of $w_1$. 
\label{Fig:uncert1}}
\end{figure}

\begin{figure}%[htbp]
\begin{center}
\begin{tabular}{cc}
%\includegraphics[scale=0.20]{w0_093-094.pdf} 
% &
%\includegraphics[scale=0.20]{w0_125-126.pdf}
% &
%\includegraphics[scale=0.20]{w0_169-170.pdf}\\

\includegraphics[scale=0.30]{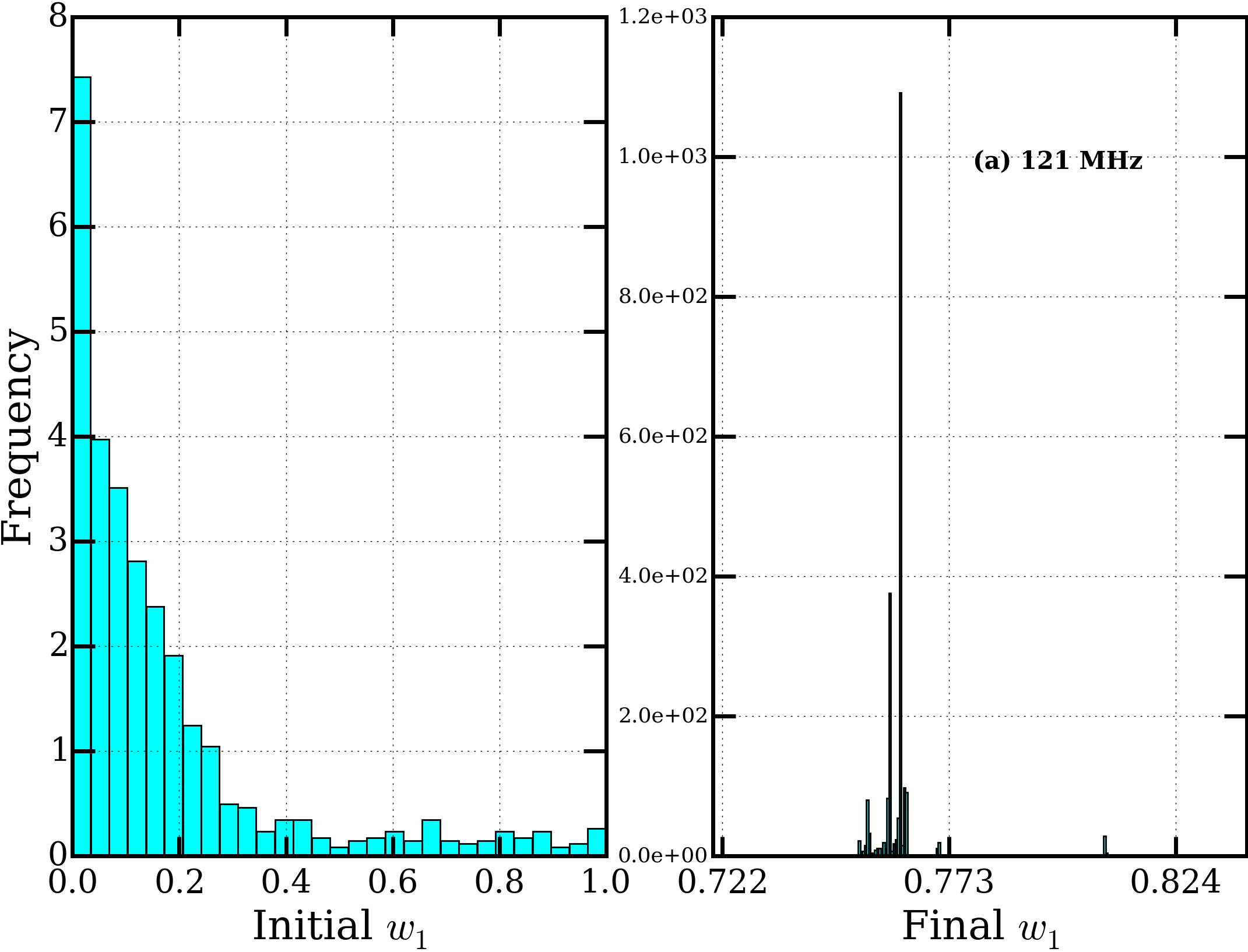}
 &
\includegraphics[scale=0.30]{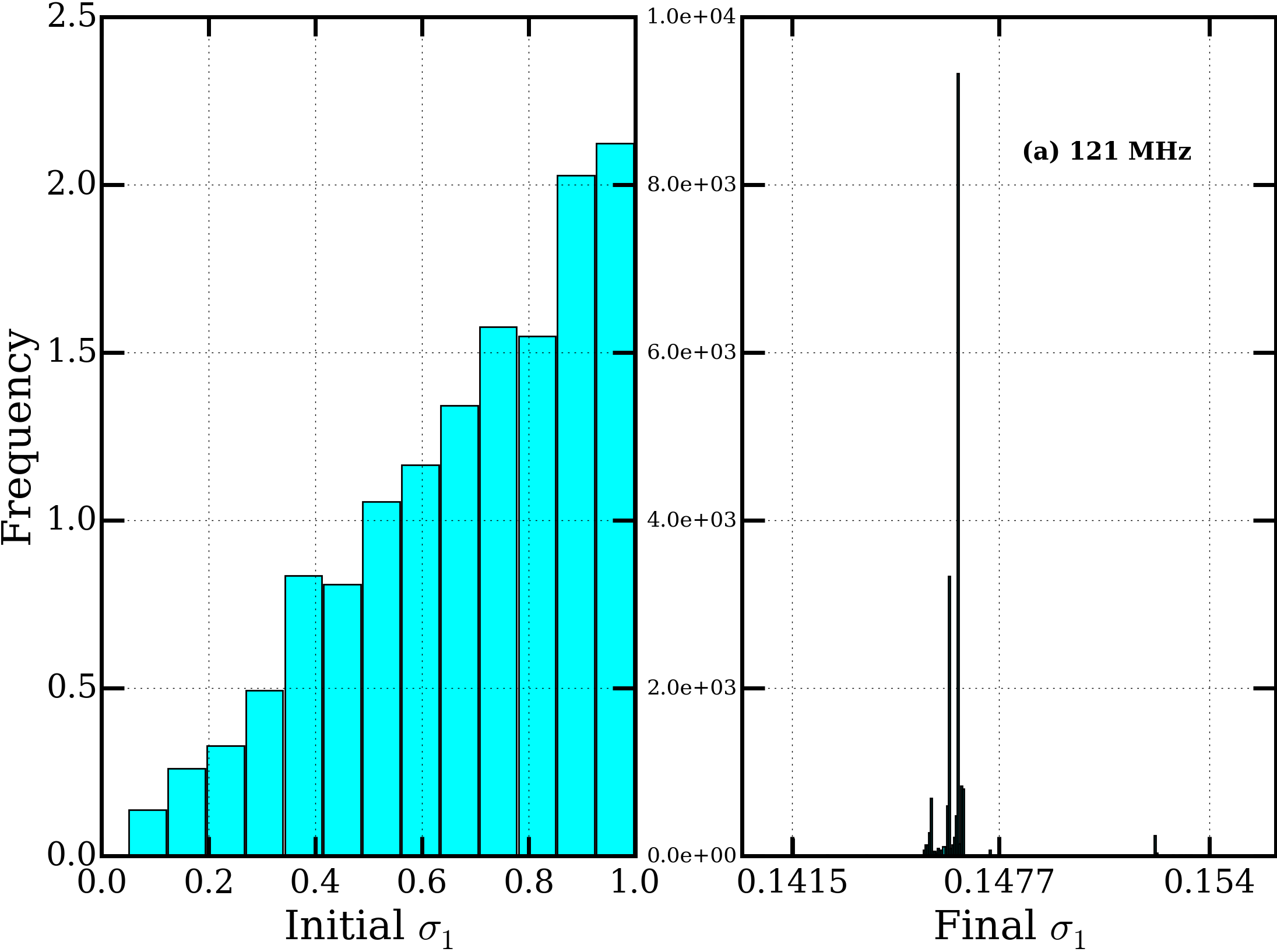}\\

\includegraphics[scale=0.30]{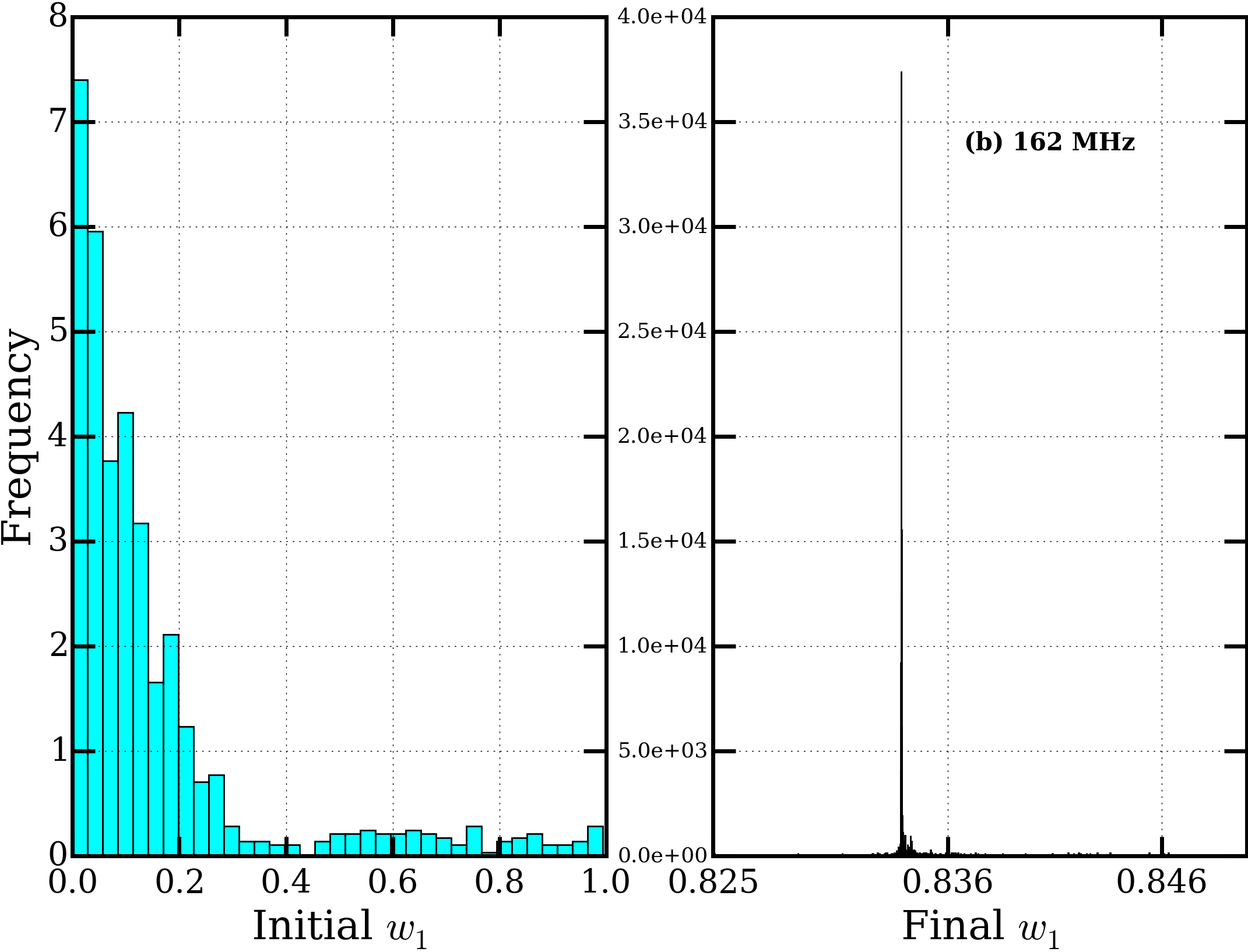}
 &
\includegraphics[scale=0.30]{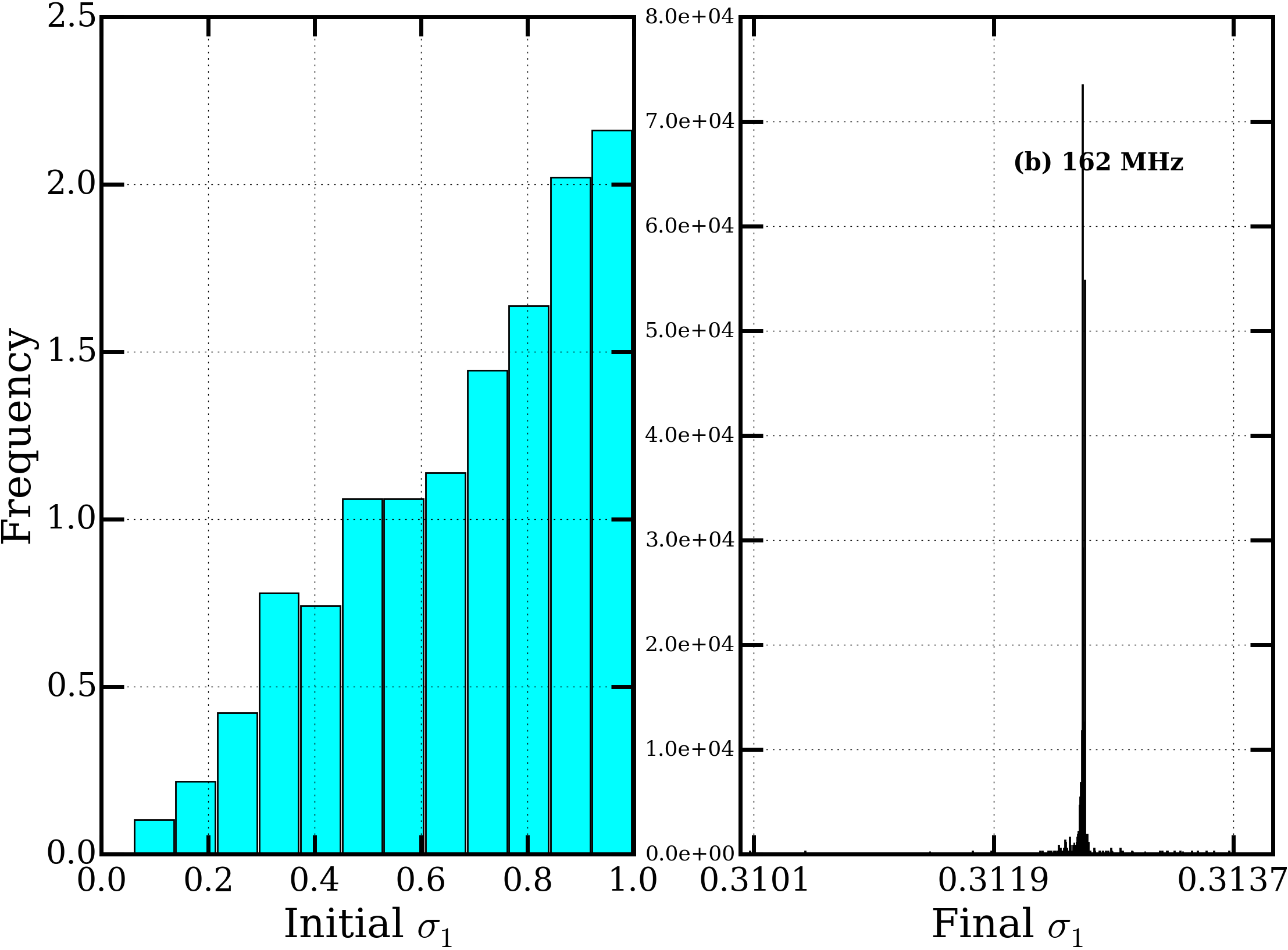}\\

\includegraphics[scale=0.30]{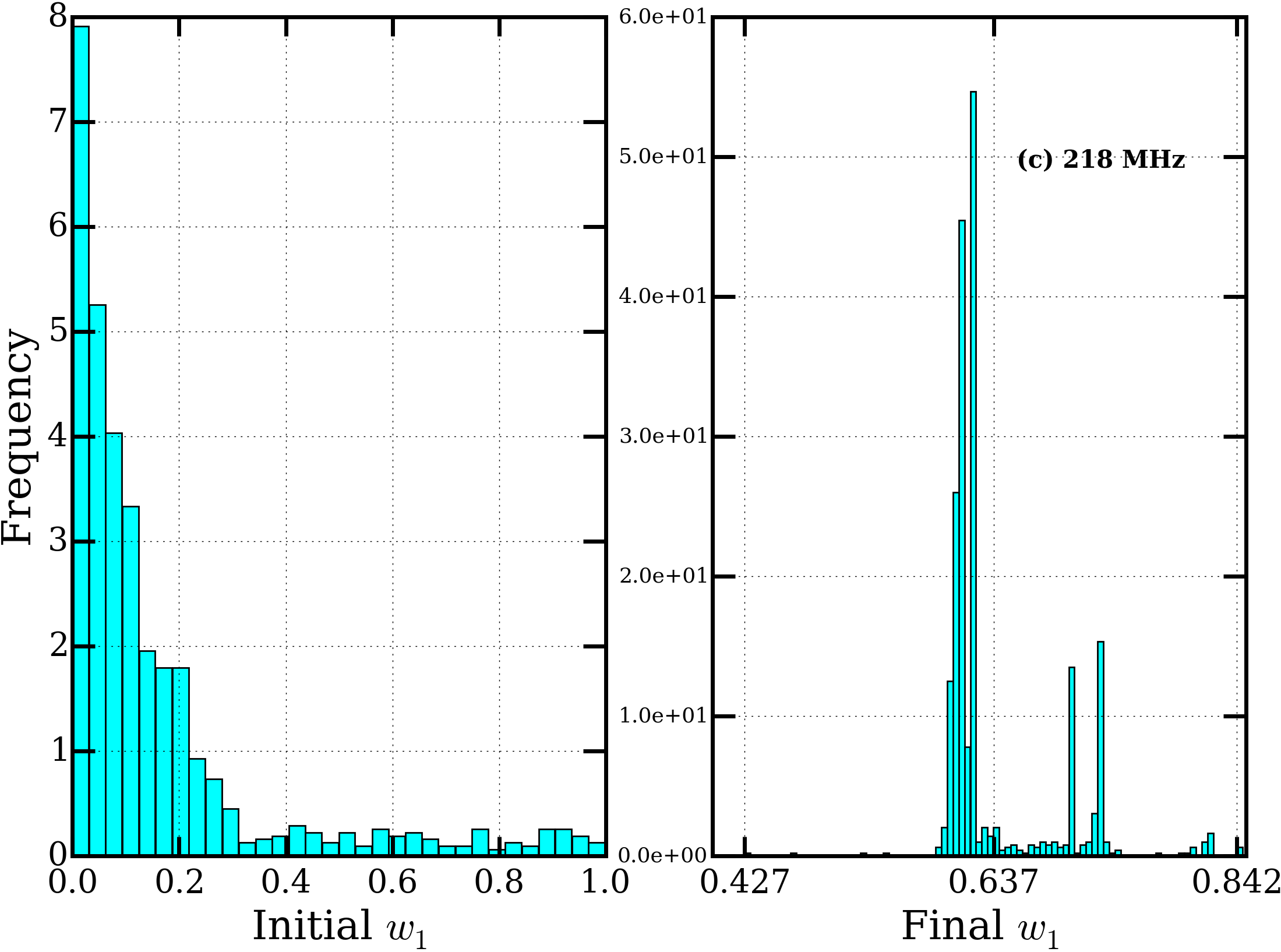}
 &
\includegraphics[scale=0.30]{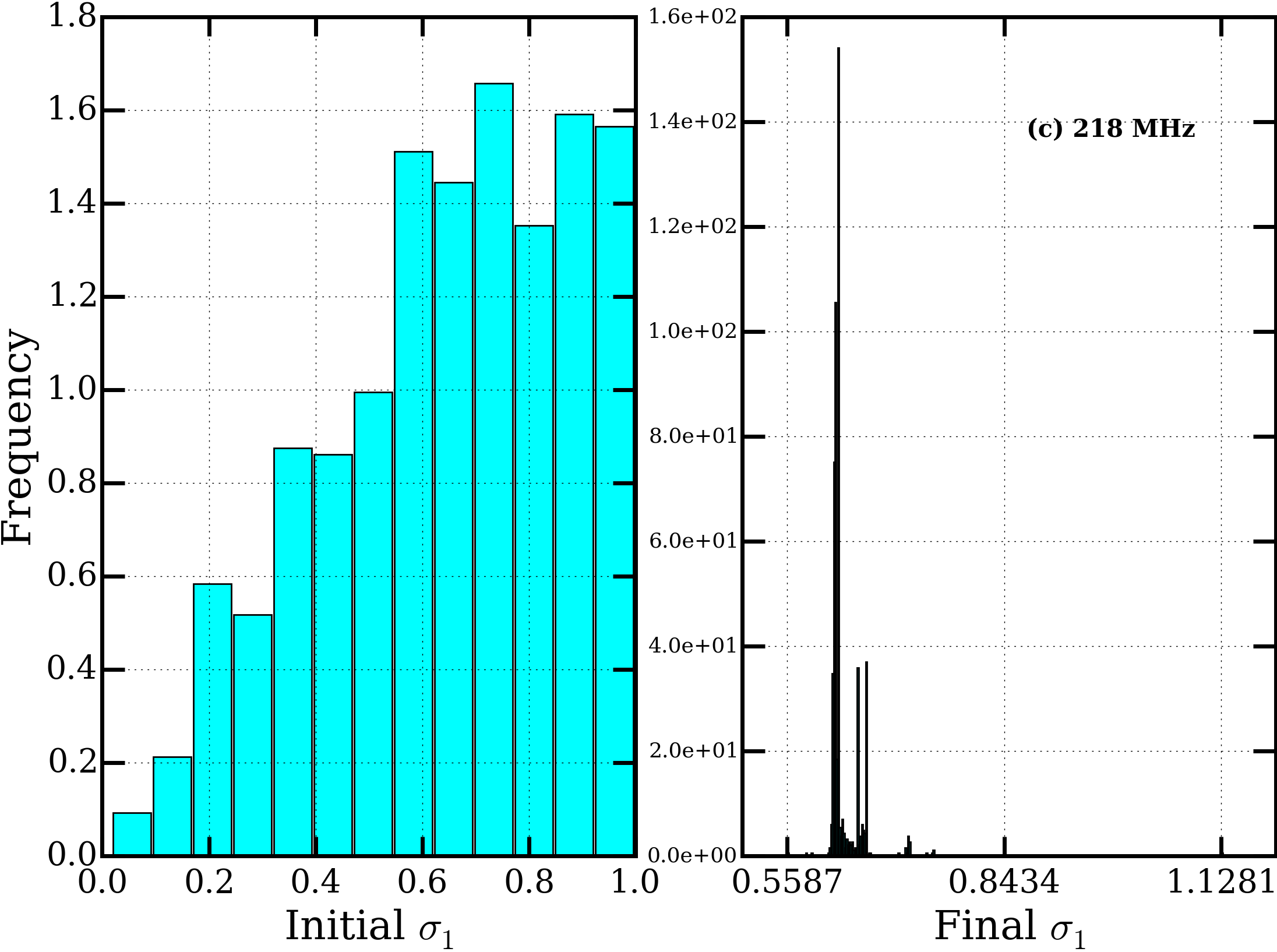}\\
\end{tabular}
\end{center}
\caption{
The left sub-panel of each of the panels shows the histogram of the initial guesses and the right sub-panel that of the final values to which GMM converged.
The left panels show plots for $w_1$s and the right panels those for $\sigma_1^2$s.
From top to bottom, the rows correspond to observing frequencies centered at 121, 162 and 218 MHz respectively.
\label{Fig:uncert2}}
\end{figure}

Both the initial guesses and the final converged values for the various parameters of the first Gaussian are shown in Fig.\ \ref{Fig:uncert1}.
This analysis was done for three of the nine frequency bands used here; those centered around 121, 162 and 218 MHz.
These frequencies were chosen to sample the entire range of variations seen in the $S_{\odot}$ distributions analyzed here (Figs.\ \ref{Fig:hist} and \ref{Fig:hists}).
The self-evident lack of clustering of colors in Fig.\ \ref{Fig:uncert1} imply absence of significant correlations between the choice of the initial seed values and the resulting final estimates.
Though the colorbars have been chosen to span the entire range of final values of $w_1$, the preponderance of symbols of very similar colors indicates that the vast majority of them lie in a very narrow range.
%%------------------------------------------------------
%%------------------------------------------------------
Figure \ref{Fig:uncert2} shows a different visualization of the same information in form of histograms of the initial and the final values of the parameters of the first Gaussian. 
%%------------------------------------------------------
%%------------------------------------------------------
All of the runs at all frequencies converge to values close to zero for the location parameter $\mu_1$, hence the corresponding histograms for $\mu_1$ are not shown. The scale parameters $\sigma_1$ also converge to very narrow ranges for the lower two frequencies, though the range increases for 218 MHz.
For the weight parameter $w_1$, the ranges spanned vary considerably, from a very tight 0.82--0.84 distribution at 162 MHz, to having a spread from 0.75--0.82 at 121 MHz and at 218 MHz a range spanning all the way from close to 0.4 to larger than 0.8.
Figure \ref{Fig:uncert2} shows that even at 218 MHz, when the range spanned by GMM $w_1$ estimates is large, the distribution is sharply peaked with only one dominant cluster the value of which is consistent with that shown in Fig.\ \ref{Fig:result_pguass}.
This implies that a small number of independent GMM runs seeded with different initial guesses are quite sufficient to arrive at robust values of the parameters of the first Gaussian with an uncertainty of $< 5\%$.

\subsection{Varying input data}
\label{Subsec:varying-input-data}
We also explored the dependence of GMM parameter estimates on the data realization, by dividing the datasets used for GMM into subsets expected to have statistically identical distributions. 
This was achieved by dividing the data into subsets with odd and even numbered time slices respectively. 
As these three datasets -- the full dataset and the ones with odd and even time slices -- are expected to have identical statistical distributions, the differences between the fitted parameters of the GMMs can provide reasonable estimates of the uncertainties associated with them. 
This analysis was carried out for data from all six baselines from each of the nine bands considered here.
In all cases, we found that the BIC criterion lead to an identical number of Gaussian components for the three datasets.
For the Gaussians with $w_i > 1$\%, the values of $w$s, $\sigma$s and $\mu$s of the individual Gaussians components were always found to be within 1\% of one another. 
This exercise thus provided an estimate of the inherent uncertainties associated with these quantities and confirmed that they are sufficiently low to not be of any practical consequence in the present context.
\section{Discussion}
\label{result}

\subsection{The slowly varying and the impulsive components}
\label{subsec:mean_S}

We separate the solar radio emission into a slowly varying and an impulsive component using a robust median filter using a window of duration 120 s.
This was motivated by a desire to separate the thermal emission from the weak impulsive emission, which must be of non-thermal origin.
As noted in Sec. \ref{Sec:solar-emission}, the slowly varying continuum component changes by $\sim$10\% in strength over time scales of many minutes.
The bandwidth of instances of impulsive emission tend to span the entire individual observing bands, each about 2 MHz wide (Fig.\ \ref{Fig:flux}).
Instances of impulsive emissions are seen across the entire spectral range covered by the observations, 100--240 MHz.
A greater density of impulsive features in a region of the time-frequency plane is usually accompanied by a local increase in the strength of the slowly varying emission.
A possible explanation for this correlation is that there exist numerous instances of impulsive emissions at time and frequency scales finer than what can be resolved by these observations (40 kHz and 0.5 s). 
These unresolved features blend into a continuum leading to the apparent increase in the strength of the slowly varying component. 
Solar type-I bursts show a similar behaviour of riding on a broadband continuum \citep{Benz1981}.
\citet{Suresh2017} found a similar trend for the emission features in $\approx$1--100 SFU range they studied using the MWA.
They also find that these features are more likely to be unresolved in time than in frequency.

The mean flux density of the slowly varying component, $S_{SV}$, is defined to be the mean of the running median filtered DS shown in the middle panel of Fig.\ \ref{Fig:flux}.
To compute the flux density of the impulsive component, $S_{Im}$, we use the observed DS from which the running median filtered DS has been subtracted, shown in the bottom panel of Fig.\ \ref{Fig:flux}.
As $\mu_1$ represents the mean of slowly varying component present in this DS, we define $S_{Im}$ as the mean flux density of its distribution after subtracting $\mu_1$, i.e.,
\begin{equation}
S_{Im} = \frac{1}{N}\sum_{i=1}^{N} (x_{i} - \mu_{1}) y_{i}\  SFU, %\ \bigg( \frac{SFU}{MHz \, sec} \bigg).
\label{Eq:eq_nth}
\end{equation}
where $x_{i}$ and $y_{i}$ are $x$ and $y$ coordinates of the histogram of flux density distribution for $i^{th}$ bin and N is the total number of histogram bins.
Table  \ref{Tab:non-thermal} lists the values of $S_{SV}$ and $S_{Im}$ for these observations. 
The rms on these numbers computed over six baselines are provided as an estimate of the associated uncertainty.

As expected, $S_{SV}$ shows a monotonic increase with frequency from 2.74 SFU at 109 MHz to 23.35 SFU at 241 MHz. 
These fluxes, however, are about 30-70\% larger than the quiet time solar fluxes obtained in \citet{Oberoi17a}. 
This is likely a consequence of the somewhat higher level of solar activity in this period when compared to that in the earlier work and is also consistent with the possible presence of unresolved non-thermal emission features mentioned earlier.
There is a general trend for $S_{Im}$ to increase with frequency as well, though it is not monotonic.
The lowest and highest values of $S_{Im}$ are 3.3 SFU at 134 MHz and 16.3 SFU at 241 MHz.
The rms on $S_{Im}$ is small and in the range 0.5--7\%.
The uncertainty on $S_{SV}$ is larger than that on $S_{Im}$.
This is expected as the flux calibration approach used here \citet{Oberoi2017a} treats all array elements as identical and does not account for tile-to-tile variations.
These variations have a larger impact on the absolute value of flux density, measured by $S_{SV}$, as compared to the differential values of flux density measured by $S_{Im}$.
We find that the $S_{SV}$ and $S_{Im}$ are rather similar in magnitude, implying that similar amounts of flux densities were radiated in the slowly varying and impulsive components during this period characterized as being of medium level of solar activity.
If we regard some fraction of $S_{SV}$ to come from unresolved impulsive non-thermal emissions, it implies that fraction of energy radiated in the non-thermal impulsive emission is even larger.

%%------------------------------------------------------
\begin{table}%[htbp]
	\centering
\caption{The continuum and impulsive fluxes, and the power-law indices for all nine frequency bands. 
The uncertainty on the fluxes correspond to 1 $\sigma$, where $\sigma$ is computed over the six baselines used here.
The uncertainty on the power-law index, $\alpha$, is computed similarly.
The last column lists $\delta \alpha$, the largest value of uncertainty on the power-law fit amongst all six baselines.}
\begin{tabular}{rrrrrr}
\tableline\tableline
\vbox{\hbox{Frequency}\hbox{(MHz)}}&\vbox{\hbox{$S_{SV}$}\hbox{(SFU)}} &\vbox{\hbox{$S_{Im}$  }\hbox{(SFU)}} & \vbox{\hbox{$w_{Im}$ }\hbox{}} &\vbox{\hbox{$\alpha$ }\hbox{}}&\vbox{\hbox{$\delta \alpha$}\hbox{}}\\
\tableline
109.0 & 2.74 $\pm$ 0.34 & 5.43 $\pm$ 0.07 & 0.25 $\pm$ 0.01 & -1.83 $\pm$ 0.08  &  0.02 \\
121.0 & 3.68 $\pm$ 1.31 & 4.62 $\pm$ 0.13 & 0.24 $\pm$ 0.00 & -1.24 $\pm$ 0.06  &  0.01 \\
134.0 & 4.84 $\pm$ 1.46 & 3.33 $\pm$ 0.13 & 0.26 $\pm$ 0.02 & -1.68 $\pm$ 0.09  &  0.03 \\
147.0 & 6.24 $\pm$ 0.74 & 5.77 $\pm$ 0.13 & 0.42 $\pm$ 0.07 & -1.47 $\pm$ 0.07  &  0.02 \\
162.0 & 8.14 $\pm$ 1.07 & 5.79 $\pm$ 0.03 & 0.17 $\pm$ 0.00 & -1.69 $\pm$ 0.06  &  0.02 \\
180.0 & 10.65 $\pm$ 1.62 & 10.44 $\pm$ 0.71 & 0.31 $\pm$ 0.03 & -1.65 $\pm$ 0.04  &  0.02 \\
198.0 & 13.54 $\pm$ 2.34 & 13.35 $\pm$ 0.89 & 0.33 $\pm$ 0.02 & -1.47 $\pm$ 0.03  &  0.02 \\
218.0 & 17.75 $\pm$ 3.02 & 12.96 $\pm$ 0.43 & 0.45 $\pm$ 0.05 & -1.61 $\pm$ 0.03  &  0.03 \\
241.0 & 23.35 $\pm$ 3.38 & 16.24 $\pm$ 0.60 & 0.28 $\pm$ 0.04 & -2.35 $\pm$ 0.07  &  0.07 \\
\tableline
\end{tabular}
\label{Tab:non-thermal}
\end{table}
%%----------------------------------------------------

The fraction of impulsive emission, $w_{Im}$, is defined to be $1-w_1$, i.e. the weight of all the data not included in the Gaussian corresponding to the slowly varying component. 
We regard $w_{Im}$ to be a measure of the prevalence of impulsive non-thermal emission, or its fractional occupancy in the frequency-time plane.
Table \ref{Tab:non-thermal} lists  $w_{Im}$ as a function of frequency along with the rms computed using $w_{Im}$ estimates from the six baselines.
$w_{Im}$ varies between 17--45\% and does not show a significant trend with frequency.
The rms on $w_{Im}$ is usually $<$10\%.
We note that the fractional occupancy of the emission with a discernible impulsive component in the DS is always rather considerable.
\citep{Thejappa1991} present a model for the type-I storms produced by the spontaneous emission by Langmuir waves due to anisotropic distribution of the electron in the closed magnetic field lines.
The randomness of the type-I bursts is believed to arise from the randomness of the density fluctuations of the suprathermal particles. 
The frequency-time occupancy of the impulsive features can potentially provide a robust observable to characterise these density fluctuations.

The impulsive coronal emissions are believed to arise from plasma emission mechanisms which give rise to emissions at the fundamental or harmonic of the local electron density and it is common practice to use coronal electron density models to ascribe a representative coronal height to the emitting regions.
Irrespective of the details of the electron density model, the absence of a well defined trend with frequency implies that the impulsive emission at different coronal heights can vary independently.
The level of activity seen at different heights is likely a reflection of the details of the coronal heights spanned by the magnetic structures along which the emitting electrons travel.
To the best of our knowledge, this is the first attempt to successfully quantify the fraction of energy being radiated in impulsive non-thermal emission and its occupancy in the time-frequency plane.

Impulsive emission at metrewavelengths is usually associated with rearrangement of magnetic flux, which in turn is associated with active regions.
This suggests the possibility that even in instances where the non-thermal emissions might remain unresolved in time and/or frequency, its morphological distribution can help investigate its association, or otherwise, with active regions and other known solar features.
As the MWA data allow imaging, these data can be used to investigate this.
This aspect, however, lies outside the scope of present work and will be undertaken in a future study.

\subsection{Flux density distribution}
\label{subsec:flux-dist}
\begin{figure}%[htbp]
\centering
\begin{tabular}{c}
\includegraphics[width=0.65\textwidth]{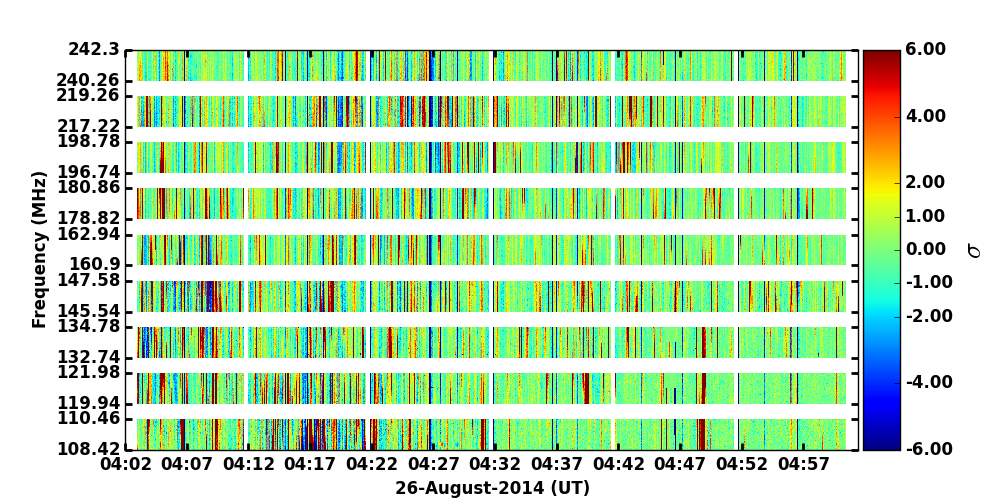}\\
\end{tabular}
\caption{The DS of $S_{Im}$ in units of $\sigma$ for 26 August 2014 from 04:02 to 05:02 UT.
\label{Fig:flux_sigma}}
\end{figure}

The most evident features from examining the histograms of $S_{Im}$ distributions shown in Figs.\ \ref{Fig:hist} and \ref{Fig:hists} are that these histograms show a clear tail at the high flux density end at all frequencies; and as we proceed from lower to higher frequencies $\sigma_1$ increases considerably.
The latter is also evident from Fig.\ \ref{Fig:result_pguass}.
Though visually instances of impulsive emission seem more numerous at higher frequencies (Fig.\ \ref{Fig:flux}), this is simply a consequence of the choice of color scale used in Fig.\ \ref{Fig:flux} and the steady increase in $\sigma_1$ with frequency.
This is demonstrated in Fig.\ \ref{Fig:flux_sigma} which shows the same DS as in the bottom most panel of Fig.\ \ref{Fig:flux} but in units of $\sigma_1$.
This provides a reliable visual measure of $w_{Im}$ consistent with the expectations based on estimates of $w_1$ (Tab. \ref{Tab:non-thermal} and Fig.\ \ref{Fig:result_pguass}).

Arranging the Gaussian components in ascending $\mu$ order and indexing them as such, one can regard $\mu_2$ as a reliable metric of the level of weakest non-thermal impulsive emission detected by this technique. 
As mentioned in Figs.\ \ref{Fig:hist} and \ref{Fig:hists}, $\mu_2$ varies from 0.18--0.93 SFU.
At the lower end of this range, these are the weakest non-thermal emissions reported in literature yet.

%----------------------------------------------------------------------

\begin{figure}
\centering
\includegraphics[width=0.8\textwidth]{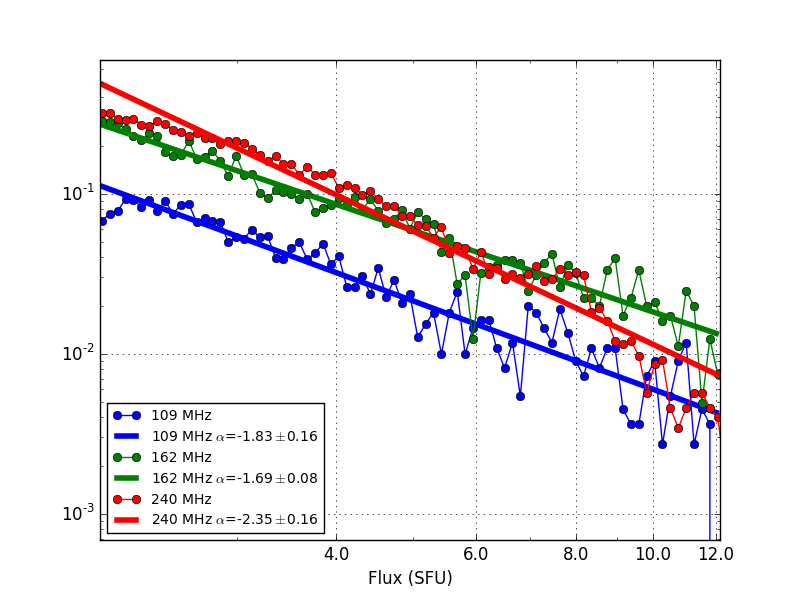}
\caption{The histogram plot showing the power-law fit to the tail of $S_{Im}$ distribution for three of the frequencies for data from one of the baselines. 
The best-fit index, $\alpha$, and the associated uncertainty are mentioned in the legend.
}
\label{Fig:power-law}
\end{figure}

In order to be energetically meaningful for nano-flare based coronal and chromospheric heating models the power law distribution of the radiated flare energies, $W$, must have an index $\alpha \leq$-2 ($dN/dW \propto W^{\alpha}$) \citep{Hudson1991}. 
It is usually very hard to estimate $W$ and $S_{Im}$ is commonly used as a proxy for $W$ \citep[e.g.][]{Saint2013}.
Therefore it is interesting to fit a power-law to the tails of these distributions.
The utility of these power-law fits comes from their ability to describe the behaviour quantitatively over large ranges.
However, we find that we are rather limited in our ability to arrive at robust fits over large spans.
At the low $S_{Im}$ end, the distribution naturally tends to flatten out and bin occupancy falls by close to two orders of magnitude as we go to higher $S_{Im}$ end, giving rise to large statistical uncertainties.
Though less than ideal, in the spirit of doing the best which these data allow, we fit power laws to these distributions in the range 2--12 SFU.
Figure \ref{Fig:power-law} shows example fits for three of the nine frequencies used here and Tab. \ref{Tab:non-thermal} lists the best-fit values of $\alpha$ for all frequencies.
The mean values of $\alpha$ presented range from -2.35 to -1.47 and the rms is computed over from all six baselines. 
The rms on $\alpha$ is in the range of few to 5\%, implying very consistent trends across baselines.
The largest value of uncertainty on the fit for a given band among all six baselines is also tabulated ($\delta \alpha$) and is usually similar to the rms on $\alpha$.
We find that the values of $\alpha$ and the associated uncertainties change significantly with small changes in the range over which the fit is done.
These estimates of $\alpha$ are therefore not robust. It will be interesting to repeat this exercise using larger data volumes which enable a more robust determination of $\alpha$.

%%----------------------------------------------------------------------
\subsection{Brightness temperature estimates}
Brightness temperature, $T_B$, is a particularly interesting diagnostic of solar non-thermal emissions. 
Over decades a vast body of information about $T_B$ associated with different kinds of non-thermal emission features has been accumulated in literature \citep[e.g.][and numerous others]{Borkowski1982,Kerdraon1983,
Melrose1989,Subramanian2004}
Most models for generation of weak non-thermal emissions suggest that they arise from localized compact sources.
In absence of imaging, based on this information we assume that the impulsive emissions being studied here come from unresolved sources. 
We use this to estimate a lower limit on their $T_B$ by assigning them an angular size of a Gaussian beam representing the snapshot monochromatic point-spread-function (PSF) for the MWA.
This approach is further bolstered by the fact in the instances when members of this team have imaged instances of weak impulsive non-thermal emissions, they have often been found to have an angular size close to the resolution limit of the instrument.
If major and minor axes of this beam are denoted by $B_{Maj}$ and $B_{Min}$ respectively, then its area is given by $\pi\ B_{Maj}\ B_{Min}$, where $B_{Maj}$ and $B_{Min}$ are expressed in radians. 
The $T_B$ and area of the MWA beam are presented in Tab. \ref{Tab:Tb}.
The PSF area decreases monotonically from 47.3 arcmin$^{2}$ to 9.7 arcmin$^{2}$, as one goes from 109 to 241 MHz.
The $T_B$ corresponding of the $S_{Im}$ among all frequency bands varies from 23 MK to 110 MK, about 1--2 orders of magnitude larger than the quiet sun $T_B$ of $\approx$ a MK.
These values lie at the low $T_B$ end of the emissions studied in most earlier reports \citep[e.g.][]{Saint2013} and are similar to the {\em microbursts} reported by \citet{Kundu1986} using the Clark Lake radioheliograph, which worked at lower frequencies (15--125 MHz) but had comparable angular resolution to the present measurements.

%In literature, the low brightness temperature (10$^{6-7}$ K) radio emission observation were reported (Kundu et al. 1986) and are called microbursts events. The timescales of the microbursts were observed to be few tens of seconds. The occurrence rate of these bursts was about several tens per day and varied depending on the active regions. The non-thermal features detected by the MWA spectrum have an upper limit of 2-11 $\times$ 10$^{11}$ K, which are close to microbursts of Kundu et al. 1986. Interestingly white et al. 1986 argues that none of the conventional mechanisms proposed for stronger type-III bursts (Scattering of Langmuir waves from Ion-polarisation clouds and quasi-linear theory) can explain the lower brightness temperature bursts ($\sim 10^{6-7}$ K). Hence there is a need to modify the type-III models or the characteristics of these bursts is different from stronger type-III and observational constraints are needed.    The MWA features provides more robust statistics for the low brightness temperature impulsive emission for the emission mechanism theories.

\begin{table}%[htbp]
\centering
\caption{The emission height, area of the PSF and brightness temperature for the nine frequency bands. The emission height is computed from newrkirk model.}
\begin{tabular}{ccc}
\tableline\tableline
\vbox{\hbox{Frequency}\hbox{(MHz)}}&\vbox{\hbox{PSF area}\hbox{(arcmin$^{2}$)}} & \vbox{\hbox{T$_{B}$}\hbox{ (K)}} \\
\tableline
109.0  &  47.3  &  3.7e+07 \\
121.0  &  38.4  &  3.2e+07 \\
134.0  &  31.3  &  2.3e+07 \\
147.0  &  26.0  &  3.9e+07 \\
162.0  &  21.4  &  4.0e+07 \\
180.0  &  17.3  &  7.3e+07 \\
198.0  &  14.3  &  9.2e+07 \\
218.0  &  11.8  &  8.9e+07 \\
241.0  &  9.7  &  1.1e+08 \\
\tableline
\end{tabular}
\label{Tab:Tb}
\end{table}

\section{Conclusion}
\label{conclusion}

We have carried out a robust statistical characterization of the weakest impulsive solar emissions reported yet at low radio frequencies.
In this work, we have shown that simple running median filtering can be used to efficiently separate slowly varying emission, believed to be of largely thermal origin, from impulsive emission, which must have a non-thermal origin.
Further, we have demonstrated that Gaussian mixture modeling used here can be used reliably to represent the observed distribution of impulsive non-thermal solar emission.
This modelling allows us to quantify, for the first time, both the amount of the energy emitted in the form of impulsive emission and also the fractional occupancy of this impulsive emission in the time-frequency plane.
In these data, taken during a period of medium level of solar activity, we show that the energy emitted in the impulsive component is very comparable to that in the slowly varying emission. 
This study also shows that at 17--45\%, the fractional occupancy of the impulsive non-thermal emission is quite substantial.
The density of these impulsive features tends to be higher in regions of increased slowly varying background emission, a behaviour analogous to type-I bursts.
It will be interesting to determine the best-fit power law slopes for the tails of these distributions and evaluate the possibility of their playing a role in coronal and chromospheric heating.
The present data is insufficient to do this reliably over a large range of flux densities.
Our approach is general enough to be useful for DS from most low radio frequency instruments, we have demonstrated it using data from the MWA.

Imaging studies to determine the spatial distribution of these instances of impulsive non-thermal emission will be crucial in establishing their association with active regions and spatio-temporal correlations with other features and phenomenon on the Sun.
These will form the subject of future studies.
We also hope that this work will generate interest in the theory and simulation community to better understand the origin and characteristics of these weak impulsive emissions.
Progress on both these fronts will be needed to evaluate their contribution towards coronal and chromospheric heating.

\acknowledgments
This scientific work makes use of the Murchison Radio-astronomy Observatory, operated by CSIRO. We acknowledge the Wajarri Yamatji people as the traditional owners of the Observatory site.  Support for the operation of the MWA is provided by the Australian Government Department of Industry and Science and Department of Education (National Collaborative Research Infrastructure Strategy: NCRIS), under a contract to Curtin University administered by Astronomy Australia Limited. We acknowledge the iVEC Petabyte Data Store and the Initiative in Innovative Computing and the CUDA Center for Excellence sponsored by NVIDIA at Harvard University.

\facility{Murchison Widefield Array}.

{\it Software:} Python ({\tt http://www.python.org}), NumPy (van der Walt et al., 2011), Scipy (van der Walt et al., 2011), Matplotlib (Hunter, 2007), Scikit-Learn (Perdegosa et al., 2011) and Scikit-Image (van der Walt et al., 2014).

\appendix

\section{Statistics of the thermal emission}

The measured interferometric cross-correlation, or visibility, is denoted by $Z$, the true visibility without thermal noise by $V$, and standard deviation of the thermal noise are related as \citep{Synthesis-Imaging-1999}:
\begin{equation}
Z^{2} = |V|^{2} + \sigma^{2}.
\label{Eq:zvs}
\end{equation}
The probability distribution of the amplitudes for the strong signal case ($V >> \sigma$) is given by,
\begin{equation}
p(Z) = \frac{1}{\sqrt{2 \pi} \sigma} \sqrt{\frac{Z}{|V|}} exp [\frac{(Z - |V|)^{2}}{2 \sigma^{2}}].
\label{Eq:probz}
\end{equation}
Under the limit $|V|>>\sigma$, the Eq. \ref{Eq:zvs} can be binomially expanded as:
\begin{equation}
<Z> = |V| (1+ \frac{\sigma^{2}}{2 |V|^{2}}).
\label{Eq:bezvs}
\end{equation}
The term $\frac{\sigma^{2}}{2 |V|^{2}}$ in the Eq. \ref{Eq:bezvs} can be computed for the strong signal case and in our case, this term is small $\sim$0.01$\%$. 
Therefore, the $\sqrt{Z/|V|}$ is $\sim 1$ in Eq. \ref{Eq:probz}, which implies that the probability distribution of the amplitudes follows a Gaussian in the strong signal regime.
\begin{equation}
p(Z) = \frac{1}{\sqrt{2 \pi} \sigma} exp [\frac{(Z - |V|)^{2}}{2 \sigma^{2}}]
%\label{Eq:probz}
\end{equation}

\bibliography{manuscript}

\end{document}